\documentclass[
	reprint,
	groupedaddress,
	amsmath,amssymb,
	prb,
	floatfix,
	superscriptaddress,
	twocolumn,
	aps,
]{revtex4-2}

\usepackage[utf8]{inputenc} %
\usepackage[T1]{fontenc}    %
\usepackage{helvet}         %
\usepackage{mathpazo}       %
\usepackage{microtype}      %
\usepackage{amsmath}
\usepackage{bm}
\usepackage{amsfonts}
\usepackage{amssymb}
\usepackage{amsthm}
\usepackage[caption=false]{subfig}
\usepackage{graphicx}
\usepackage{hyperref}
\usepackage{cleveref}
\usepackage[T1]{fontenc}
\usepackage{multirow}
\usepackage{lipsum} %
\usepackage{color,soul}
\usepackage[export]{adjustbox}
\usepackage{natbib}

\usepackage{todonotes}

\usepackage{xr}
\begin{document}

\title{Chemical site bases and average-atom potentials for the atomic cluster expansion}

\author{Lorenzo Piersante}
\affiliation{Laboratory of Materials Design and Simulation (MADES), Institute of Materials, \'{E}cole Polytechnique F\'{e}d\'{e}rale de Lausanne}

\author{Anirudh Raju Natarajan}
\email{anirudh.natarajan@epfl.ch}
\affiliation{Laboratory of Materials Design and Simulation (MADES), Institute of Materials, \'{E}cole Polytechnique F\'{e}d\'{e}rale de Lausanne}

\date{\today}

\begin{abstract}
    Interatomic potentials are central tools in the atomistic modeling of materials. The atomic cluster expansion (ACE) parameterizes such potentials from \textit{ab initio} data, conventionally encoding the chemical degrees of freedom with a one-hot representation that yields chemically stratified models. The alternative chemical representations used in on-lattice configurational cluster expansions have not been assessed for interatomic potentials. Here we revisit the multicomponent ACE for an arbitrary chemical site basis. We then establish an exact analytical mapping between a fitted linear ACE and the average-atom potential that describes a perfectly random alloy. We benchmark potentials built on the occupational, Chebyshev, and conventional ACE bases against solute binding and vacancy formation energies in Mg--Nd, and against the mixing enthalpies of the Mo--Nb and Cr--W solid solutions. When training data are scarce, the occupational basis converges fastest and offers the best control over targeted material properties, while the conventional and Chebyshev bases face challenges in reproducing these properties. The occupational basis likewise yields the most reliable average-atom description of disordered alloy thermodynamics. In the large-data limit the three bases perform identically. The chemical basis is therefore a design choice that governs data efficiency. Its explicit treatment opens a route to average-atom potentials for the thermodynamic, mechanical, and kinetic properties of concentrated alloys.
\end{abstract}

\maketitle

\section{Introduction}
Interatomic potentials are essential tools for investigating atomic-scale processes \cite{deringer_machine_2019, friederich_machine-learned_2021} across systems ranging from battery materials \cite{guo_intercalation_2023, deringer_modelling_2020} to metallic alloys \cite{eyert_machine-learned_2023, rahman_review_2025}. Universal models \cite{mazitov_pet-mad_2025, marchand_foundation_2025, lysogorskiy_graph_2026} now predict energies and forces across most of the periodic table, but reaching the accuracy needed for a specific system often requires substantial fine-tuning \cite{radova_fine-tuning_2025,wong_bias_2026}. Bespoke potentials sit at the opposite end of the spectrum, with hyperparameters and training data tuned to the accuracy--efficiency tradeoff of the modeling task at hand \cite{toit_hyperparameter_2024, piersante_machine_2026}. We focus on the linear atomic cluster expansion (ACE), which decomposes the total energy into site energies, $E = \sum_{i=1}^N E_i$. Each site energy is expanded over a complete, hierarchical cluster basis \cite{drautz_atomic_2019, drautz_atomic_2020}. The resulting models are systematically expandable, interpretable, and computationally efficient \cite{lysogorskiy_performant_2021, bochkarev_efficient_2022}. They have been used to study structural and carbon-based materials \cite{ibrahim_atomic_2023, qamar_atomic_2023}, liquids \cite{ibrahim_atomic_2026}, and oxygen diffusion \cite{bienvenu_development_2025}.

For multicomponent systems, the ACE is normally formulated with a one-hot representation \cite{drautz_atomic_2020} for each chemical species. This representation is chemically stratified, in the sense that it yields a separate site-energy model for each species. An existing ACE potential can therefore be extended to chemically more complex alloys without modifying the basis functions of the species already present. The close relationship between the ACE and the on-lattice configurational cluster expansion \cite{sanchez_generalized_1984} indicates that alternative chemical bases are equally admissible \cite{sanchez_cluster_2010}. Because these bases are related by linear transformations, the choice is immaterial for a complete expansion. The choice matters in practice, where the expansion is truncated at a finite body order and the coefficients are regularized on a finite training set. The consequences of the chemical basis choice for learning specific material properties have not been assessed \cite{dusson_atomic_2022}.

Bespoke potentials resolve the interactions of individual chemical configurations, but many applications instead require the properties of a chemically disordered phase. Average-atom potentials describe a random alloy in which every site is occupied by an effective species set by the alloy composition. Such an alloy is perfectly random, or mean-field, in the sense that it carries no short-range order. These average-atom potentials give access to the mixing enthalpy and to average atomistic properties without sampling chemical configurations. They play an established role in the study of concentrated and high-entropy alloys, where they underpin solute-strengthening theory and the simulation of averaged dislocation and grain-boundary properties \cite{smith_application_1989, varvenne_average-atom_2016, hodapp_exact_2025}.

Computing the properties of the mean-field alloy with an on-lattice configurational cluster expansion is well established. Replacing the site basis functions by their composition-weighted expectation values yields the energy of a perfectly random alloy \cite{sanchez_cluster_2010}. Disordered-alloy energetics can also be obtained by direct sampling of random configurations, or with approximants such as special quasirandom structures \cite{zunger_special_1990}. The same substitution is not straightforward for a body-ordered interatomic potential, because the linear-scaling formulation introduces self-interactions that a naive averaging treats incorrectly. Hodapp \cite{hodapp_exact_2025} has proposed a formalism for linear machine-learning potentials written in multilinear form, but, as pointed out by the author, a shortcoming of that approach is the reduced efficiency of the average potential. A route for converting a linear ACE into an average-atom potential, in the symmetrized basis used in practice and for an arbitrary chemical site basis, is still missing. Making the chemical degrees of freedom explicit is the first step towards a rigorous derivation. 

Here, we generalize the multicomponent ACE to an arbitrary chemical site basis. The comparison holds the continuous degrees of freedom fixed and varies only the chemical representation. Alongside the conventional ACE basis we consider the Chebyshev and occupational bases familiar from on-lattice cluster expansions. Because the chemical degrees of freedom are explicit, a fitted ACE can be mapped analytically onto its average-atom counterpart. We derive this mapping together with the exact treatment of the self-interactions, so that all standard training techniques for machine-learning interatomic potentials transfer directly to average-atom models. The computational cost of the average potential is then equivalent to an ACE potential for a single component system. The three bases are assessed on global error metrics and on solute binding and vacancy formation energies in dilute Mg--Nd. Mixing enthalpies of the concentrated Mo--Nb and Cr--W solid solutions are evaluated directly from the average-atom potential. The chemical basis governs both data efficiency and the accuracy of targeted properties when training data are scarce, although the three bases become equivalent once the training set is large. Our study focuses on metallic alloys, but the framework is transferable to other classes of materials, such as semiconductor and thermoelectric alloys.

\section{Chemical site bases for the atomic cluster expansion} \label{sect:formulation}
The multicomponent ACE describes the site energy $E_i$ of atom $i$ in terms of its local environment. Consider atom $i$ and the $N$ atoms in its neighborhood, each site occupied by one of $M$ possible chemical species. The site energy depends on the $N$ displacement vectors from the center atom, $\boldsymbol{r}_{i}^{N} = (\boldsymbol{r}_{i1}, \boldsymbol{r}_{i2}, \dots, \boldsymbol{r}_{iN})$. It also depends on the species of the center atom and its neighbors, collected in the tuple $\boldsymbol{\mu}_i^{N+1} = (\mu_i, \mu_1, \dots, \mu_N)$. A superscript counts the entries of a tuple, so $\boldsymbol{\mu}_i^{N+1}$ carries one species for the center atom and one for each neighbor. We write the site energy function as $E_i(\boldsymbol{\mu}_i^{N+1}, \boldsymbol{r}_i^N)$.

The continuous degrees of freedom of $E_i$ require a complete basis of infinitely many bond functions $\phi_\nu (\boldsymbol{r}_{ij})$, with $\nu = 0, 1, 2, \dots$. The chemical degrees of freedom instead require only $M$ site basis functions $\sigma_{\eta}(\mu_j)$, with $\eta = 0, 1, 2, \dots, M-1$. These functions take the species $\mu_j$ of site $j$ as their argument.

A cluster $\alpha = (\boldsymbol{\mu}_i^{K+1}, \boldsymbol{r}_i^K)$ comprises the center atom $i$ and $K$ of its neighbors. Each cluster gives rise to multicomponent cluster functions of the form:
\begin{gather}
    \label{eq:cluster_functions}
    \Phi_{\boldsymbol{\eta}\boldsymbol{\nu}\alpha} = \Phi_{\boldsymbol{\eta}\alpha}\Phi_{\boldsymbol{\nu}\alpha} \\
    \Phi_{\boldsymbol{\eta}\alpha} = \sigma_{\eta_0}(\mu_i)\prod_{k=1}^K \sigma_{\eta_k}(\mu_k)\\
    \Phi_{\boldsymbol{\nu}\alpha} = \prod_{k=1}^K \phi_{\nu_k} (\boldsymbol{r}_{ik})
\end{gather}
where $\boldsymbol{\eta} = (\eta_0, \eta_1, \dots, \eta_K)$ is a tuple of site basis function indices and $\boldsymbol{\nu} = (\nu_1, \nu_2, \dots, \nu_K)$ is a tuple of bond function indices. The index $\eta_0$ labels the site basis function of the center atom. The cluster function $\Phi_{\boldsymbol{\eta}\alpha}$ carries the chemical degrees of freedom of the $K+1$ atoms, while $\Phi_{\boldsymbol{\nu}\alpha}$ carries the continuous degrees of freedom of the $K$ bonds.

Setting $\sigma_0(\mu_j) = 1$ and $\phi_0(\boldsymbol{r}_{ij}) = 1$ makes the multicomponent cluster basis hierarchical. The site energy is then given by:
\begin{equation}
    E_i(\boldsymbol{\mu}_i^{N+1}, \boldsymbol{r}_i^N) = \sum_{\alpha}\sum_{\boldsymbol{\eta}}\sum_{\boldsymbol{\nu}} J_{\boldsymbol{\eta}\boldsymbol{\nu}} \Phi_{\boldsymbol{\eta}\boldsymbol{\nu}\alpha}
    \label{eq:general_mACE}
\end{equation}
where the sum runs over all clusters $\alpha$ contained in the environment of atom $i$, and $J_{\boldsymbol{\eta}\boldsymbol{\nu}}$ are expansion coefficients. These coefficients carry no explicit cluster index because the dependence on the cluster is implicit in the length and entries of $\boldsymbol{\eta}$ and $\boldsymbol{\nu}$ \cite{drautz_atomic_2019, drautz_atomic_2020}.

In practice, \cref{eq:general_mACE} is not used directly because the sum over clusters scales polynomially with the number of neighbors. The cluster functions are also not guaranteed to be invariant under permutation, rotation, and inversion of the environment \cite{drautz_atomic_2019}. Linear scaling and permutation symmetry are recovered by introducing multicomponent atomic density functions, defined as sums over the neighbors of atom $i$:
\begin{equation}
    A_{\eta \nu} = \sum_{j=1}^N \sigma_\eta(\mu_j) \phi_\nu(\boldsymbol{r}_{ij})
    \label{eq:neighbourhood_expansion}
\end{equation}
The bond functions are products of a radial function, $R_{n}(r_{ij})$, and a spherical harmonic, $Y_l^m(\hat{\boldsymbol{r}}_{ij})$:
\begin{equation}
    \phi_{nlm}(\boldsymbol{r}_{ij}) = R_{n}(r_{ij})Y_l^m(\hat{\boldsymbol{r}}_{ij})
\end{equation}
where $\nu$ is a tuple of radial and angular indices, $(nlm)$.

The $A$-basis collects the multicomponent cluster functions for a cluster with $K$ neighbors, each being a product of the site basis function of the center atom and $K$ atomic densities:
\begin{equation}
    A_{\boldsymbol{\eta}\boldsymbol{\nu}} = \sigma_{\eta_0}(\mu_i) \prod_{k=1}^{K} A_{\eta_k \nu_k}
\end{equation}
Invariance under rotation and inversion requires linear combinations of several $A_{\boldsymbol{\eta}\boldsymbol{\nu}}$ such that the total angular momentum of the spherical harmonics vanishes and the sum of the $l$ quantum numbers is even. The Clebsch-Gordan iteration outlined by Drautz \cite{drautz_atomic_2019, drautz_atomic_2020} generates the $B$-basis of symmetrized multicomponent cluster functions:
\begin{equation}
    B_{\boldsymbol{\eta}\boldsymbol{\nu}} = \sigma_{\eta_0}(\mu_i)\sum_{\boldsymbol{m}} \left (\begin{matrix} \boldsymbol{l} \\ \boldsymbol{L} \end{matrix} 0 \right ) \prod_{k=1}^{K} A_{\eta_k \nu_k}
\end{equation}
where $\left (\begin{matrix} \boldsymbol{l} \\ \boldsymbol{L} \end{matrix} 0 \right )$ is a generalized Clebsch-Gordan coefficient that combines the $l$ and intermediate $L$ quantum numbers to give zero total angular momentum \cite{drautz_atomic_2020, dusson_atomic_2022, goff_permutation-adapted_2024}. The symmetrized site energy is then given by:
\begin{equation}
    E_i(\boldsymbol{\mu}_i^{N+1}, \boldsymbol{r}_i^N) = \sum_{\boldsymbol{\eta}}\sum_{\boldsymbol{\nu}} c_{\boldsymbol{\eta}\boldsymbol{\nu}} B_{\boldsymbol{\eta}\boldsymbol{\nu}}
    \label{eq:ACE_site_energy}
\end{equation}
where the length of the index tuples fixes the body order of the cluster, and $c_{\boldsymbol{\eta}\boldsymbol{\nu}}$ are fitting coefficients. This expansion is formally infinite, and it is truncated in practice so that a finite number of coefficients can be parameterized.

\subsection{Chemical site basis functions} \label{sect:chemical_bases}
The chemical occupation variable $\mu$ of a system with $M$ species takes values from 0 to $M-1$. Describing the possible occupations of each site requires $M$ linearly independent site basis functions:
\begin{equation}
    \boldsymbol{\sigma}(\mu) = \begin{pmatrix}
        \sigma_0(\mu) \\ \sigma_1(\mu) \\ \vdots \\ \sigma_{M-1}(\mu)
    \end{pmatrix}
\end{equation}
Two site bases are common in on-lattice configurational cluster expansions: the occupational basis and the Chebyshev basis, the latter also known as the spin basis \cite{sanchez_generalized_1984}. The occupational basis gives rise to a generalized lattice gas model, whereas the Chebyshev basis produces a generalized Ising model \cite{barroso-luque_cluster_2024}. The examples below consider a binary system of A and B atoms.

The occupational and Chebyshev bases for a binary system are:
\begin{equation}
    \boldsymbol{\sigma}_{\mathrm{Occ}} =
    \begin{pmatrix}
        1 \\  \delta_{\mu B}
    \end{pmatrix}
    \quad \quad
    \boldsymbol{\sigma}_{\mathrm{Cheby}} =
    \begin{pmatrix}
    1 \\ \frac{2 \delta_{\mu B} - 1}{\sqrt{2}}
    \end{pmatrix}
\end{equation}

The Chebyshev basis can be written in terms of Chebyshev polynomials and is orthogonal, whereas the occupational basis is not. Both are hierarchical because $\sigma_0(\mu) = 1$.

The conventional ACE basis builds chemically stratified expansions in which species can be added or removed without modifying the basis functions of the remaining species. Cartesian unit vectors achieve this, but they are not hierarchical. Hierarchy is restored by enlarging the chemical space with a hypothetical \emph{vacuum} species that carries the constant site basis function and is never sampled \cite{drautz_atomic_2020}. The resulting chemical basis is:
\begin{equation}
    \boldsymbol{\sigma}_{\mathrm{ACE}} =
    \begin{pmatrix}
      1 \\ \delta_{\mu A} \\ \delta_{\mu B}
    \end{pmatrix}
\end{equation}
Within the ACE site basis, every expansion coefficient $c_{\boldsymbol{\eta}\boldsymbol{\nu}}$ whose tuple $\boldsymbol{\eta}$ contains a vacuum species label is set to zero.

\begin{figure*}[!ht]
    \centering
    \includegraphics[width=0.9\linewidth]{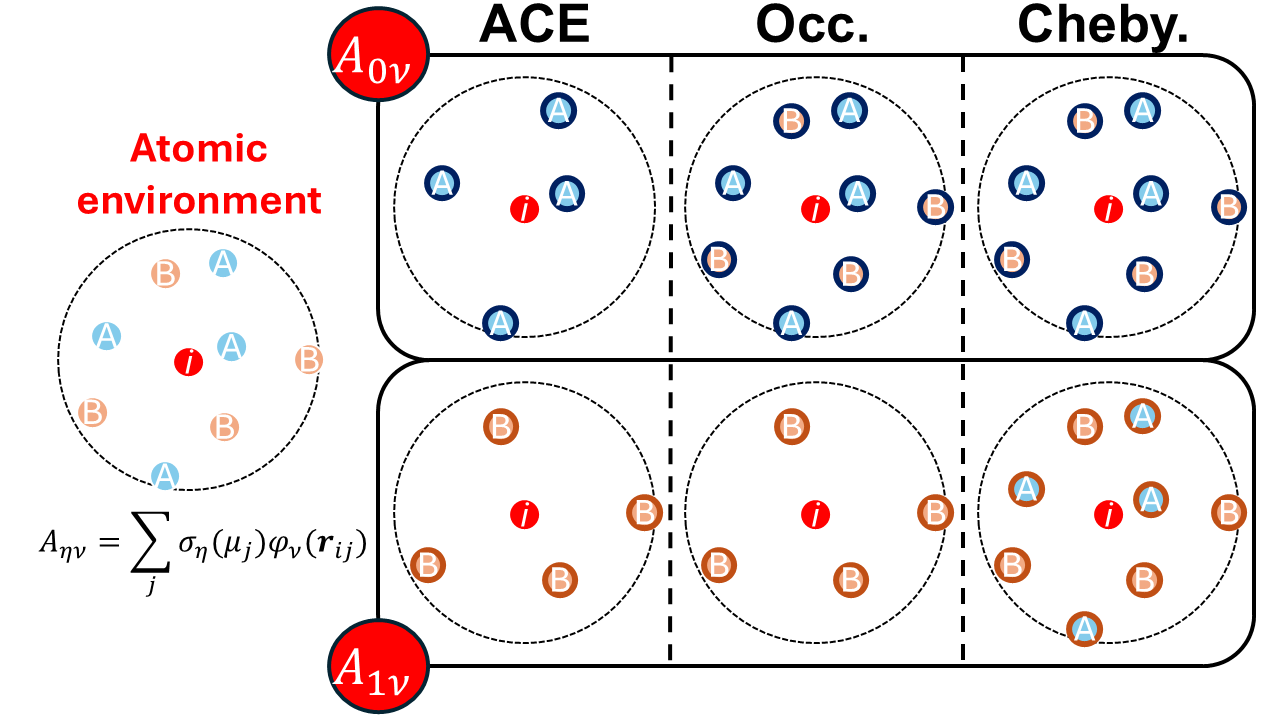}
    \caption{Atomic densities resolved by the functions $A_{\eta \nu}$ (\cref{eq:neighbourhood_expansion}) according to the conventional ACE, occupational, and Chebyshev bases. The reference environment is a binary system of A and B atoms centered on atom $i$. The upper row shows the density associated with the site basis function $\sigma_0$, and the lower row the density corresponding to $\sigma_1$. The schematic omits the vacuum species of the conventional ACE basis.}
    \label{fig:density_schematic}
\end{figure*}

Different chemical site bases resolve different atomic densities around atom $i$. \Cref{fig:density_schematic} illustrates these densities for the conventional ACE, occupational, and Chebyshev bases. The ACE basis resolves two densities, one for each atomic species, whereas the hierarchical bases resolve a background density and a second, chemically sensitive density. The background density responds only to structural features because $\sigma_0$ is independent of the chemical occupation. The second density encodes the distribution of B atoms for the occupational basis, and a combination of A and B atoms for the Chebyshev basis.

Any two site bases are related by a linear transformation $\Lambda$ that maps $\sigma_\eta(\mu)$ onto a different site basis $\tau_\kappa(\mu)$:
\begin{equation}
    \tau_\kappa(\mu) = \sum_{\eta = 0}^{M-1} \Lambda_{\kappa\eta}\sigma_\eta(\mu)
    \label{eq:site_transformation}
\end{equation}
A multicomponent cluster function $\Phi_{\boldsymbol{\eta}\boldsymbol{\nu}\alpha}$ transforms in the same way into a cluster function $\tilde{\Phi}_{\boldsymbol{\kappa}\boldsymbol{\nu}\alpha}$ built on the new site basis:
\begin{equation}
    \tilde{\Phi}_{\boldsymbol{\kappa}\boldsymbol{\nu}\alpha} = \sum_{\eta_0}\sum_{\eta_1}\cdots\sum_{\eta_K} \Lambda_{\kappa_0 \eta_0}\Lambda_{\kappa_1 \eta_1}\cdots\Lambda_{\kappa_K \eta_K} \Phi_{\boldsymbol{\eta}\boldsymbol{\nu}\alpha}
    \label{eq:cluster_site_transformation}
\end{equation}
\Cref{eq:site_transformation,eq:cluster_site_transformation} show that atomic cluster expansions built on different chemical bases map onto each other, so the choice of basis is immaterial for a complete expansion. For pair potentials, the transformation between the expansion coefficients of two chemical bases follows trivially.

\section{Average-atom atomic cluster expansion} \label{sect:average_atom}
\subsection{Disordered-phase descriptors} \label{sect:disordered_descriptors}
A linear ACE with explicit chemical site basis functions can be converted into an average-atom interatomic potential that computes the energies, forces, and stresses of the mean-field alloy. The conversion replaces the chemical site basis functions of \cref{eq:ACE_site_energy} by functions of the alloy composition. Because the site energy is linear in the descriptors $B_{\boldsymbol{\eta}\boldsymbol{\nu}}$, the coefficients $c_{\boldsymbol{\eta}\boldsymbol{\nu}}$ carry over unchanged. The task therefore reduces to evaluating the descriptors in the disordered phase. We seek a factorized form in which a purely \emph{geometric descriptor}, evaluated as though every site carried the same species, multiplies a scalar \emph{chemical factor} fixed by the alloy composition. The coefficients of the parent ACE then weight this product. As in \cref{sect:chemical_bases}, we present the derivation for a binary A-B system.

In a random alloy the occupation probability of each site is fixed by the composition, and the occupations of different sites are independent. We denote by $\langle \cdots \rangle$ an average over the chemical configurations of such an alloy. Every site basis function then takes the value $\langle \sigma_{\eta} \rangle$, so that each site carries the same average atom, set by the composition \cite{zunger_special_1990, sanchez_cluster_2010}. The expectation values of the occupational, Chebyshev, and conventional ACE site bases in a binary alloy are:
\begin{equation}
    \begin{gathered}
    \langle \boldsymbol{\sigma}_{\mathrm{Occ}} \rangle = \begin{pmatrix} 1 \\ x \end{pmatrix} \\
    \langle \boldsymbol{\sigma}_{\mathrm{Cheby}} \rangle = \begin{pmatrix} 1 \\ \tfrac{2 x - 1}{\sqrt{2}} \end{pmatrix}\\
    \langle \boldsymbol{\sigma}_{\mathrm{ACE}} \rangle = \begin{pmatrix} 1 - x \\ x \end{pmatrix}
    \end{gathered}
    \label{eq:site_basis_averages}
\end{equation}
where the vacuum species is omitted from the ACE basis and $x$ denotes the atomic fraction of species B. Averaging \cref{eq:ACE_site_energy} over the chemical configurations yields the average-atom site energy $\overline{E}_i$:
\begin{equation}
    \overline{E}_i(x, \boldsymbol{r}_i^N) = \sum_{\boldsymbol{\eta}}\sum_{\boldsymbol{\nu}} c_{\boldsymbol{\eta}\boldsymbol{\nu}} \langle B_{\boldsymbol{\eta}\boldsymbol{\nu}} \rangle
    \label{eq:average_atom_site_energy}
\end{equation}
The averaging replaces the dependence on the chemical occupation by a dependence on the composition. Constructing an average-atom potential then amounts to expressing the disordered descriptors $\langle B_{\boldsymbol{\eta}\boldsymbol{\nu}} \rangle$ as geometric descriptors that an ACE implementation can evaluate rapidly, each multiplied by a scalar chemical factor that is fixed once the alloy composition is chosen.

The pair cluster illustrates how the disordered descriptors separate into a geometric descriptor and a chemical factor. The pair descriptor, $B_{\eta_0 \eta_1 n 0} = \sigma_{\eta_0}(\mu_i) \sum_{j=1}^{N} \sigma_{\eta_1}(\mu_j) R_n(r_{ij}) Y_0^0$, carries two site basis functions evaluated on the distinct sites $i$ and $j$. The occupations of these two sites are independent, so the configurational average factorizes into a product of expectation values:
\begin{equation}
    \langle B_{\eta_0 \eta_1 n 0} \rangle = \langle \sigma_{\eta_0} \rangle \langle \sigma_{\eta_1} \rangle  B_{n 0}
  \label{eq:average_atom_pair_ace}
\end{equation}
where $B_{n 0} = \sum_{j=1}^{N} R_n(r_{ij}) Y_0^0$ is the geometric pair descriptor, evaluated on the structure as though it contained a single species. This is an ordinary pair descriptor of the parent expansion, which an existing ACE implementation already evaluates. The product $\langle \sigma_{\eta_0} \rangle \langle \sigma_{\eta_1} \rangle$ is the chemical factor, which \cref{eq:site_basis_averages} supplies for each chemical site basis at a given alloy composition.

\begin{figure}[!ht]
    \includegraphics[width=\linewidth]{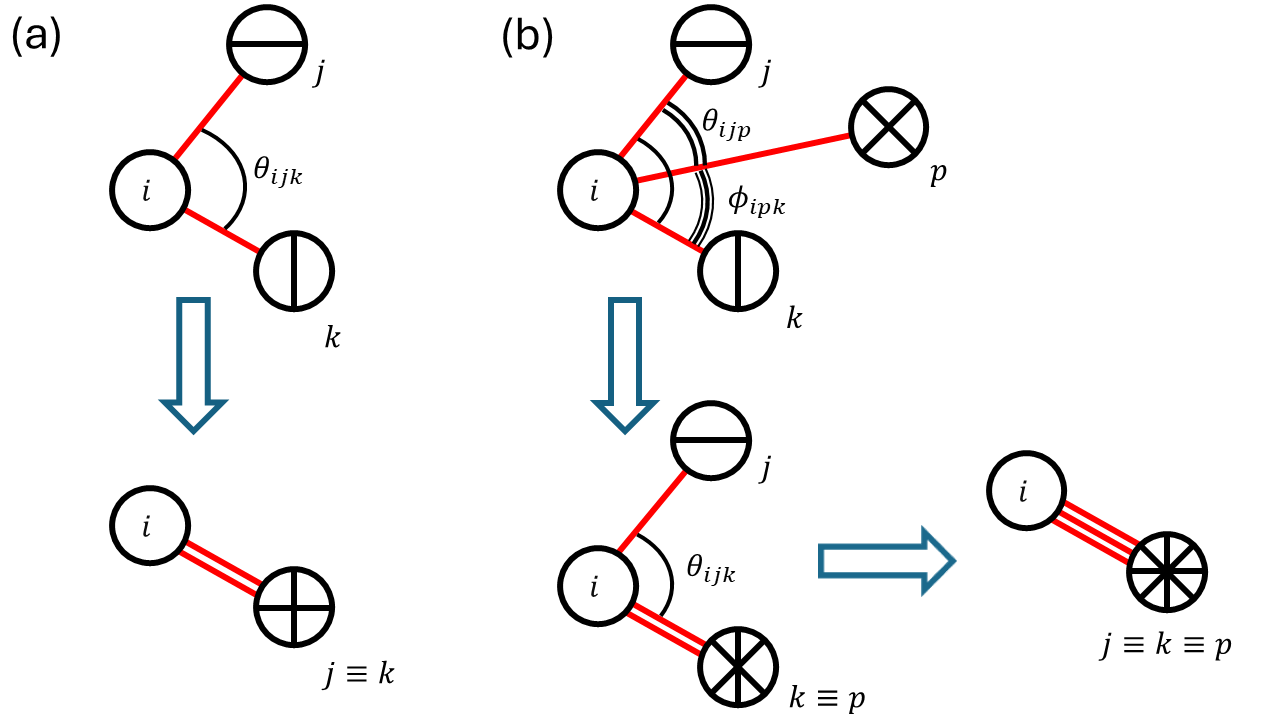}
    \caption{Self-interacting clusters generated by the chemical averaging of triplet (a) and quadruplet (b) $B$-basis functions. $\theta$ and $\phi$ mark bond and dihedral angles. Double and triple lines mark bonds formed by the superposition of two and three coincident bonds.}
    \label{fig:self_interaction_schematic}
\end{figure}

In contrast to pair clusters, triplet clusters complicate this factorization and require separate treatment of the self-interactions. The triplet descriptor is:
\begin{equation}
    B_{\eta_0 \eta_1 \eta_2 n_1 n_2 l} = \sigma_{\eta_0}(\mu_i) \sum_{j=1}^N \sum_{k = 1}^N \sigma_{\eta_1}(\mu_j) \sigma_{\eta_2}(\mu_k) \Phi_{n_1 n_2 l}
    \label{eq:triplet_basis_fn}
\end{equation}
where $\Phi_{n_1 n_2 l}(r_{ij}, r_{ik}, \theta_{ijk})$ is the symmetrized product of two radial basis functions and two spherical harmonics. The double sum includes the diagonal $j = k$, in which both neighbor site basis functions are evaluated on the same atom. These diagonal terms are the self-interactions that accompany the linear-scaling formulation of the ACE \cite{drautz_atomic_2019,ho_atomic_2024}. Geometrically, they are triplets whose bond angle has collapsed to zero so that neighbors $j$ and $k$ coincide, leaving a pair cluster carrying two superposed bonds, as shown in \cref{fig:self_interaction_schematic}(a). Self-interactions are harmless in the multicomponent ACE because lower-order contributions cancel the self-interactions generated at higher body order \cite{drautz_atomic_2019}.

Separating the diagonal $j = k$ from the remaining terms gives:
\begin{equation}
    \begin{aligned}
        B_{\eta_0 \eta_1 \eta_2 n_1 n_2 l} = \sigma_{\eta_0}(\mu_i) \sum_{j} \sum_{k \neq j} \sigma_{\eta_1}(\mu_j) \sigma_{\eta_2}(\mu_k) \Phi_{n_1 n_2 l} \\
        + \sigma_{\eta_0}(\mu_i) \sum_{j} \sigma_{\eta_1}(\mu_j) \sigma_{\eta_2}(\mu_j) \Phi^{\mathrm{self}}_{n_1 n_2 l}
    \end{aligned}
    \label{eq:factorised_triplet_basis_fn}
\end{equation}
where $\Phi^{\mathrm{self}}_{n_1 n_2 l} = \Phi_{n_1 n_2 l}(r_{ij}, r_{ij}, 0)$. The first term involves three distinct sites, so its average factorizes into a product of expectation values. The second term evaluates both neighbor site basis functions on the same neighbor, so its average retains the joint expectation value $\langle \sigma_{\eta_1} \sigma_{\eta_2} \rangle$:
\begin{equation}
    \begin{aligned}
        \langle B_{\eta_0 \eta_1 \eta_2 n_1 n_2 l} \rangle = \langle \sigma_{\eta_0} \rangle \langle \sigma_{\eta_1} \rangle \langle \sigma_{\eta_2} \rangle \sum_{j} \sum_{k \neq j} \Phi_{n_1 n_2 l} \\
        + \langle \sigma_{\eta_0} \rangle \langle \sigma_{\eta_1} \sigma_{\eta_2} \rangle \sum_{j} \Phi^{\mathrm{self}}_{n_1 n_2 l}
    \end{aligned}
    \label{eq:mf_triplet_basis_fn}
\end{equation}
Substituting $\langle \sigma_{\eta} \rangle$ for $\sigma_{\eta}$ everywhere in \cref{eq:triplet_basis_fn} would replace $\langle \sigma_{\eta_1} \sigma_{\eta_2} \rangle$ by $\langle \sigma_{\eta_1} \rangle \langle \sigma_{\eta_2} \rangle$ and misrepresent the self-interactions \cite{hodapp_exact_2025}.

\begin{table}[!ht]
    \centering
    \begin{tabular}{|c|c|c|c|} \hline
        Basis & $\mathrm{Cov}(\sigma_0, \sigma_0)$ & $\mathrm{Cov}(\sigma_0, \sigma_1)$ & $\mathrm{Cov}(\sigma_1, \sigma_1)$ \\ \hline
        Occupational & 0 & 0 & $x (1 -x)$\\
        Chebyshev & 0 & 0 & $2x(1-x)$\\
        ACE & $x (1 -x)$ & $-x (1 -x)$ & $x (1 -x)$ \\ \hline
    \end{tabular}
    \caption{Covariances of the site basis functions, $\mathrm{Cov}(\sigma_{\eta_1}, \sigma_{\eta_2}) = \langle \sigma_{\eta_1} \sigma_{\eta_2} \rangle - \langle \sigma_{\eta_1} \rangle \langle \sigma_{\eta_2} \rangle$, for the occupational, Chebyshev, and conventional ACE bases in a binary A--B alloy at atomic fraction $x$ of species B. These covariances weight the self-interaction descriptors in \cref{eq:unconstrained_mf_triplet_basis_fn}.}
    \label{tab:binary_covariances}
\end{table}

\Cref{eq:mf_triplet_basis_fn} is exact, but the restricted double sum is impractical to evaluate. Excluding the diagonal requires explicit bookkeeping over neighbor pairs, which the atomic densities of \cref{eq:neighbourhood_expansion} do not provide. Adding and subtracting the diagonal, evaluated with factorized expectation values, restores the unrestricted sums:
\begin{equation}
    \begin{aligned}
    \langle B_{\eta_0 \eta_1 \eta_2 n_1 n_2 l} \rangle = \langle \sigma_{\eta_0} \rangle \langle \sigma_{\eta_1} \rangle \langle \sigma_{\eta_2} \rangle B_{n_1 n_2 l} \\ + \langle \sigma_{\eta_0}\rangle \mathrm{Cov}(\sigma_{\eta_1}, \sigma_{\eta_2}) B^{\mathrm{self}}_{(n_1 n_2) l}
    \end{aligned}
    \label{eq:unconstrained_mf_triplet_basis_fn}
\end{equation}

Two distinct geometric descriptors emerge. The first is the conventional geometric triplet descriptor $B_{n_1 n_2 l} = \sum_{j} \sum_{k} \Phi_{n_1 n_2 l}$, recovered from the parent ACE by stripping away the site basis functions. The second is a self-interaction descriptor:
\begin{align}
  B^{\mathrm{self}}_{(n_1 n_2) l} &= \sum_{j} \Phi^{\mathrm{self}}_{n_1 n_2 l} \nonumber\\
                                  &= \sum_{j} R_{n_1}(r_{ij}) R_{n_2}(r_{ij}) \sum_{m} C^{00}_{l m l (-m)} Y_{l}^{m}(\hat{\boldsymbol{r}}_{ij}) Y_{l}^{-m}(\hat{\boldsymbol{r}}_{ij})
  \label{eq:triplet_self_interaction}
\end{align}
where $C^{00}_{l m l (-m)}$ denotes a Clebsch-Gordan coefficient that combines two spherical harmonics to produce an invariant feature. This descriptor is a single sum over neighbors in which both radial functions and spherical harmonics are evaluated on the same bond. The alloy composition enters only through the chemical prefactors, which differ between the two terms. A product of single-site expectation values weights the conventional descriptor $B_{n_1 n_2 l}$, whereas the covariance $\mathrm{Cov}(\sigma_{\eta_1}, \sigma_{\eta_2}) = \langle \sigma_{\eta_1} \sigma_{\eta_2}\rangle - \langle \sigma_{\eta_1} \rangle \langle \sigma_{\eta_2}\rangle $ weights the self-interaction descriptor. \Cref{tab:binary_covariances} lists these covariances for the three chemical bases in a binary alloy. Both descriptors remain sums over the neighborhood of atom $i$, so the average-atom expansion retains the linear scaling of the ACE. Evaluating the energy of the disordered phase therefore hinges on one quantity, $B^{\mathrm{self}}_{(n_1 n_2) l}$, that a conventional ACE implementation does not directly supply.

\begin{figure*}[!ht]
    \centering
    \includegraphics[width=0.8\linewidth]{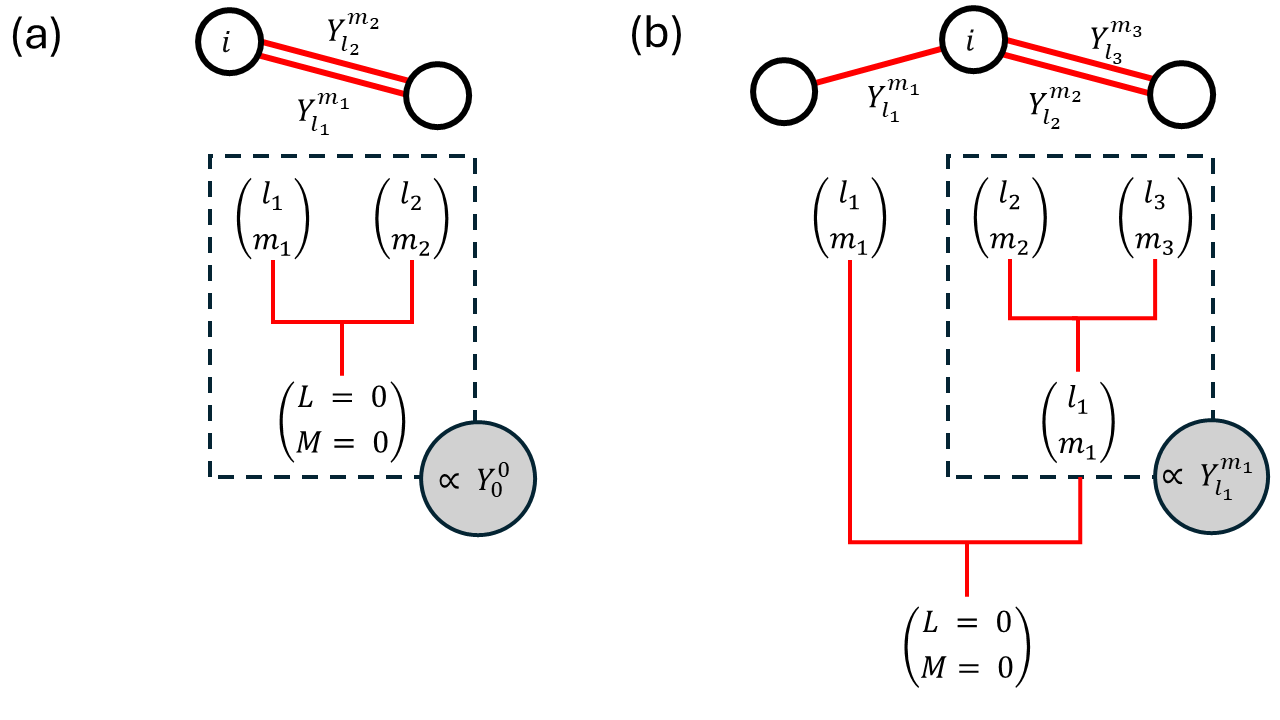}
    \caption{Reduction of the Clebsch-Gordan iteration for self-interacting clusters. The Clebsch-Gordan contraction of two spherical harmonics evaluated on the same bond simplifies according to \cref{eq:inverse_cg_series}. (a) Reduction for a triplet self-interaction. (b) Reduction for a quadruplet self-interaction involving two coincident bonds.}
    \label{fig:cg_iteration}
\end{figure*}

\subsection{Self-interaction descriptors} \label{sect:self_interaction}
The self-interaction descriptors of triplet and quadruplet clusters reduce to ordinary ACE descriptors of lower body order. Each is evaluated on a radial basis augmented with products of the original radial functions, and weighted by a scalar prefactor that depends only on the angular indices. An existing ACE implementation can therefore evaluate them with its usual Clebsch-Gordan machinery. These closed forms follow from the properties of spherical harmonics. The self-interacting clusters themselves arise from the bond-coincidence patterns of the parent $B$-basis function. As summarized in \cref{fig:self_interaction_schematic}, the self-interaction term of a triplet yields pair-like descriptors, whereas that of a quadruplet yields both triplet-like and pair-like descriptors.

Repeated application of the \emph{inverse Clebsch-Gordan series} yields these closed forms. The series contracts two spherical harmonics evaluated on the same bond \cite{varshalovich_quantum_1988}:
\begin{equation}
    \sum_{m_1, m_2} C_{l_1 m_1 l_2 m_2}^{L M} Y_{l_1}^{m_1}(\hat{\boldsymbol{r}}) Y_{l_2}^{m_2}(\hat{\boldsymbol{r}}) = N_{l_1 l_2}^{L} C_{l_1 0 l_2 0}^{L 0} Y_L^M(\hat{\boldsymbol{r}})
    \label{eq:inverse_cg_series}
\end{equation}
where $C_{l_1 m_1 l_2 m_2}^{L M}$ is a Clebsch-Gordan coefficient and $N_{l_1 l_2}^{L}$ is a scalar prefactor fixed by the angular indices, given in \cref{supp-sect:self_interaction} of the Supporting Information. \Cref{fig:cg_iteration}(a) shows the Clebsch-Gordan iteration acting on the product of spherical harmonics inside \cref{eq:triplet_self_interaction}, in which the two harmonics couple to a total angular momentum $L = 0$. A single application of \cref{eq:inverse_cg_series} to that inner sum gives:
\begin{equation}
    B_{(n_1 n_2) l}^{\text{self}} =  N_{l l}^{0} C_{l 0 l 0}^{0 0} B_{(n_1 n_2)0}
    \label{eq:triplets_two_bonds}
\end{equation}
where $B_{(n_1 n_2)0} = \sum_j R_{n_1 n_2}(r_{ij})Y_0^0$ is a pair descriptor built on the product radial basis $R_{n_1 n_2}(r) = R_{n_1}(r)R_{n_2}(r)$. The triplet self-interaction therefore reduces to a pair descriptor weighted by two scalars that depend only on $l$. In what follows, bracketed $n$ and $l$ labels denote radial and angular functions evaluated on the same bond.

Quadruplet self-interactions follow in the same way. A quadruplet carries three bonds, so either two of them coincide or all three do, as shown in \cref{fig:self_interaction_schematic}(b). These two patterns give the averaged quadruplet descriptor two types of self-interaction terms, in contrast to the single term of \cref{eq:unconstrained_mf_triplet_basis_fn}.

We begin with two coincident bonds, which arise in three ways. The bonds of the $(l_2, m_2)$ and $(l_3, m_3)$ spherical harmonics may coincide while the $(l_1, m_1)$ harmonic carries a distinct bond, as illustrated in \cref{fig:cg_iteration}(b). The two remaining possibilities pair $(l_1, m_1)$ with $(l_2, m_2)$, and $(l_1, m_1)$ with $(l_3, m_3)$. The coincident pair undergoes the first Clebsch-Gordan contraction, producing an intermediate quantum number that couples with the remaining spherical harmonic. Zero total angular momentum requires this intermediate number to equal $l_1$, the index of the harmonic on the free bond. Both harmonics of this first contraction are evaluated on the same bond, so \cref{eq:inverse_cg_series} reduces it to a single spherical harmonic multiplied by a scalar prefactor where the upper index $L$ equals $l_1$, while the lower indices are $l_2$ and $l_3$. The second layer of the iteration produces the triplet-like angular dependence of degree $l_1$, and the resulting self-interaction descriptor is:
\begin{equation}
    B_{n_1 (n_2 n_3) l_1 (l_2 l_3)}^{\text{self}} = N_{l_2 l_3}^{l_1} C_{l_2 0 l_3 0}^{l_1 0} B_{n_1 (n_2 n_3) l_1}
    \label{eq:quardruplets_two_bonds}
\end{equation}
The other two descriptors of this kind follow from the permutation symmetry of the invariant combination of three spherical harmonics \cite{edmonds_angular_1960}.

When all three bonds coincide, two successive applications of \cref{eq:inverse_cg_series} yield the self-interaction descriptor:
\begin{equation}
    B_{(n_1 n_2 n_3) (l_1 l_2 l_3)}^{\text{self}} = N_{l_1 l_1}^{0} C_{l_1 0 l_1 0}^{0 0} N_{l_2 l_3}^{l_1} C_{l_2 0 l_3 0}^{l_1 0} B_{(n_1 n_2 n_3) 0}
    \label{eq:quardruplets_three_bonds}
\end{equation}
where $B_{(n_1 n_2 n_3) 0} = \sum_j R_{n_1 n_2 n_3}(r_{ij})Y_0^0$ is a pair descriptor built on the triple product radial basis $R_{n_1 n_2 n_3}(r) = R_{n_1}(r)R_{n_2}(r)R_{n_3}(r)$.

Collecting the geometric descriptors and chemical factors gives the averaged quadruplet descriptor, derived in detail in \cref{supp-sect:quadruplet_descriptors} of the Supporting Information:
\begin{equation}
    \begin{aligned}
    \langle B_{\boldsymbol{\eta} \boldsymbol{n} \boldsymbol{l}} \rangle = \langle \sigma_{\eta_0} \rangle \Big [ \langle \sigma_{\eta_1} \rangle \langle \sigma_{\eta_2} \rangle \langle \sigma_{\eta_3} \rangle B_{\boldsymbol{n} \boldsymbol{l}} \\
    + \sum_{a} \sum_{(b, c)} \langle \sigma_{\eta_a} \rangle \text{Cov}(\sigma_{\eta_b}, \sigma_{\eta_c}) B^{\text{self}}_{n_a (n_b n_c) l_a (l_b l_c)} \\
    + \text{Cov}(\sigma_{\eta_1}, \sigma_{\eta_2}, \sigma_{\eta_3})  B^{\text{self}}_{(n_1 n_2 n_3) (l_1 l_2 l_3)} \Big ]
    \end{aligned}
    \label{eq:quadruplet_averaged}
\end{equation}
where the double sum runs over the three ways of leaving one bond $a$ free and collapsing the pair $(b, c)$. The symbol $\text{Cov}(\sigma_{\eta_1}, \sigma_{\eta_2}, \sigma_{\eta_3})$ denotes the joint central moment $\left \langle \prod_a (\sigma_{\eta_a} - \langle \sigma_{\eta_a} \rangle) \right \rangle$. Each of the three terms pairs a geometric descriptor with a chemical factor. The first term carries the ordinary quadruplet descriptor $B_{\boldsymbol{n} \boldsymbol{l}}$, a four-body interaction weighted by a product of three expectation values. The second carries $B^{\text{self}}_{n_a (n_b n_c) l_a (l_b l_c)}$, a three-body interaction on a radial basis augmented with the product $R_{n_b n_c}$, weighted by the covariance of the two site basis functions that share a bond. The third carries $B^{\text{self}}_{(n_1 n_2 n_3) (l_1 l_2 l_3)}$, a two-body interaction on the product basis $R_{n_1 n_2 n_3}$, weighted by the joint central moment of all three. Collapsing bonds therefore lowers the body order of the geometric descriptor and raises the order of the chemical moment that weights it.

\Cref{eq:triplets_two_bonds,eq:quardruplets_two_bonds,eq:quardruplets_three_bonds} provide a practical route to evaluating the self-interaction descriptors. We augment the radial basis with the products $R_{n_b n_c}$ and $R_{n_1 n_2 n_3}$, evaluate ordinary triplet and pair descriptors, and apply the scalar prefactors afterwards. The average-atom potential therefore preserves the linear scaling of the ACE. \Cref{supp-sect:self_interaction} of the Supporting Information collects analytical expressions for every self-interaction descriptor generated by triplet and quadruplet clusters.

\subsection{Average-atom site energy} \label{sect:average_atom_energy}
Combining the results of \cref{sect:disordered_descriptors,sect:self_interaction} gives the average-atom site energy up to quadruplet descriptors:
\begin{equation}
    \overline{E}_i(x, \boldsymbol{r}_i^N) = \sum_{\boldsymbol{\nu}} \overline{c}_{\boldsymbol{\nu}}(x) B_{\boldsymbol{\nu}} + \sum_{\tilde{\boldsymbol{\nu}}} \overline{c}_{\tilde{\boldsymbol{\nu}}}^{\text{self}}(x) B_{\tilde{\boldsymbol{\nu}}}^{\text{self}}
    \label{eq:average_atom_site_energy_2}
\end{equation}
where the overlined coefficients are composition-weighted sums over the chemical degrees of freedom, and the label $\tilde{\boldsymbol{\nu}}$ keeps track of the self-interacting $n$ and $l$ indices. Both sets of geometric descriptors are evaluated as though every site were occupied by the same average atom, so the coefficients contain the entire dependence on composition.

Converting a fitted ACE potential into its average-atom counterpart therefore requires only the composition-dependent interaction coefficients. The coefficients $\overline{c}_{\boldsymbol{\nu}}(x)$ take the same form at every body order:
\begin{equation}
    \overline{c}_{\boldsymbol{\nu}}(x) = \sum_{\boldsymbol{\eta}} \left ( \prod_{k = 0}^K \langle \sigma_{\eta_k} \rangle \right )c_{\boldsymbol{\eta} \boldsymbol{\nu}}
    \label{eq:standard_coefficients}
\end{equation}
where the product runs over the center atom and the $K$ neighbors of the cluster. By contrast, the self-interaction coefficients take a different form at each body order. The triplet self-interaction coefficient, for instance, is:
\begin{equation}
     \overline{c}^{\text{self}}_{n_1 n_2 l}(x) = \sum_{\eta_0, \eta_1, \eta_2} \langle \sigma_{\eta_0} \rangle \text{Cov}(\sigma_{\eta_1}, \sigma_{\eta_2}) c_{\eta_0 \eta_1 \eta_2 n_1 n_2 l}
     \label{eq:self-interaction_coefficients}
\end{equation}
Evaluating \cref{eq:standard_coefficients,eq:self-interaction_coefficients} from the coefficients of a fitted ACE supplies the interaction parameters of the average-atom potential.

\section{Methods}
Linear ACE models were fitted to \textit{ab initio} datasets with each of the three chemical site bases. The comparison covers both the quality of the interatomic potential and the thermodynamic properties predicted by the average-atom potentials derived from these models. Energies and forces of the structures in all datasets were computed with density functional theory (DFT) as implemented in the \textit{Vienna Ab initio Simulation Package} (VASP, version 6.4.1) \cite{kresse_ab_1993, kresse_ab_1994, kresse_efficient_1996, kresse_ultrasoft_1999}. The calculations used the Perdew-Burke-Ernzerhof exchange-correlation functional within the generalized gradient approximation \cite{perdew_generalized_1996, perdew_generalized_1997} and projector-augmented wave pseudopotentials. Calculation settings, including the valence electron configuration of each element, the plane-wave cutoff, the $k$-point mesh spacing, and the second-order Methfessel-Paxton smearing width, are summarized in \cref{supp-sect:dft_calculations} of the Supporting Information. Part of the data were taken from earlier work \cite{piersante_machine_2026, lee_modeling_2026}.

The potentials were trained by linear regression. Potentials containing only pair descriptors were fitted by ordinary least squares. All other fits used Ridge regression or the least absolute shrinkage and selection operator (Lasso). The $L_1$ or $L_2$ regularization strength was selected by leave-one-out cross-validation, which is appropriate for the small datasets considered here.

In the absence of regularization terms, the loss function is:
\begin{equation}
    \mathcal{L} = \sum_{s=1}^S(E_s^{\mathrm{DFT}} - \hat{E}_s)^2 + \sum_{s=1}^S \sum_{i=1}^{N_s}||\boldsymbol{F}_{si}^{\mathrm{DFT}} - \hat{\boldsymbol{F}}_{si}||^2
\end{equation}
where the hat indicates the model prediction, $S$ is the number of structures, and $N_s$ is the number of atoms in structure $s$. We fitted to total energies for the point-defect properties, and to per-atom energies in all other cases. No weighting scheme was applied, so that models built on different site bases could be compared on equal footing. Because the radial basis functions go smoothly to zero at the cutoff radius, the DFT energies were referenced to isolated dimers separated by the cutoff distance.

Our implementation of the ACE relies on the \verb|metatensor| software ecosystem \cite{bigi_fast_2023, bigi_metatensor_2026}. We evaluated the atomic densities $A_{\eta \nu}$ and the self-interaction descriptors with a customized version of the \verb|torch-spex| package. The $B$-basis functions were obtained with the \verb|DensityCorrelations| routine of the \verb|featomic| package, using the basis-set sizes listed in \cref{supp-sect:potential_config} of the Supporting Information. The site energy was implemented as a \verb|pytorch| model built on routines from the \verb|metatrain| and \verb|metatomic| packages. The resulting model supports automatic differentiation, which yields the forces and the derivatives of the $B$-basis functions.
 
\section{Results} \label{sect:results}
We assessed how the accuracy of a fitted potential changes when the chemical site basis is varied among the conventional ACE, Chebyshev, and occupational bases. We also examined how the chemical basis affects the mixing enthalpy evaluated with the corresponding average-atom potential. \Cref{sect:error_metrics_point_defects} describes the dilute limit of Mg--Nd, from global error metrics to solute binding and vacancy formation energies, while \cref{sect:random_alloy} describes the mixing enthalpies of the concentrated Mo--Nb and Cr--W solid solutions. \Cref{supp-sect:datasets,supp-sect:potential_config} of the Supporting Information give the composition of each dataset and the interatomic potential settings.

\begin{figure}[!ht]
    \centering
    \subfloat[][\label{fig:basis_functions}]{\includegraphics[height=0.5\linewidth]{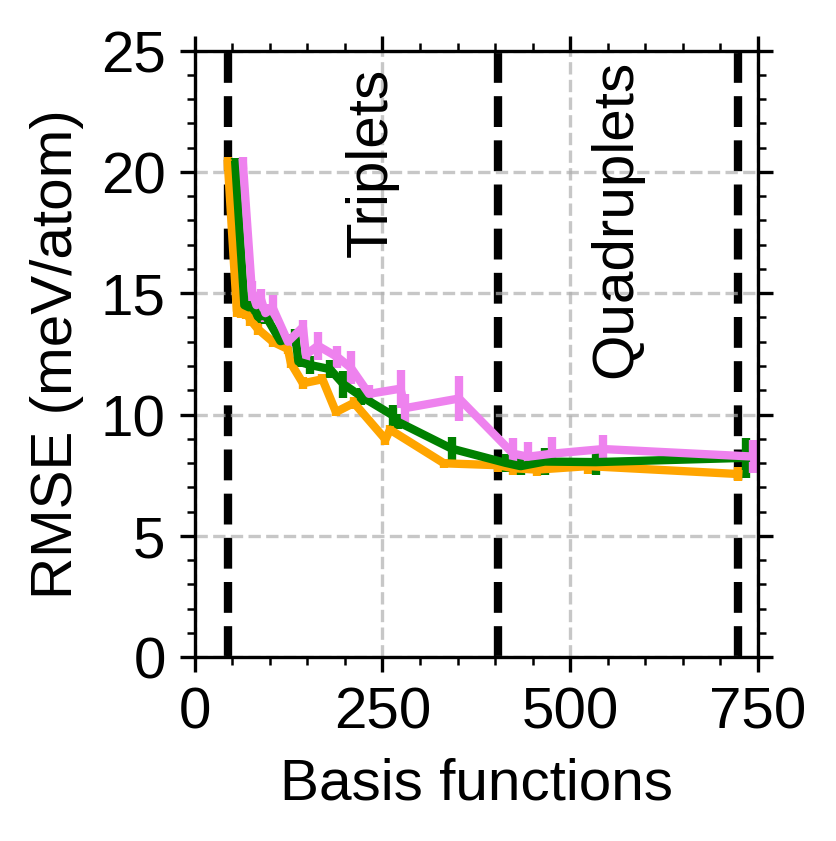}}
    \subfloat[][\label{fig:learning_curve_E}]{\includegraphics[height=0.5\linewidth]{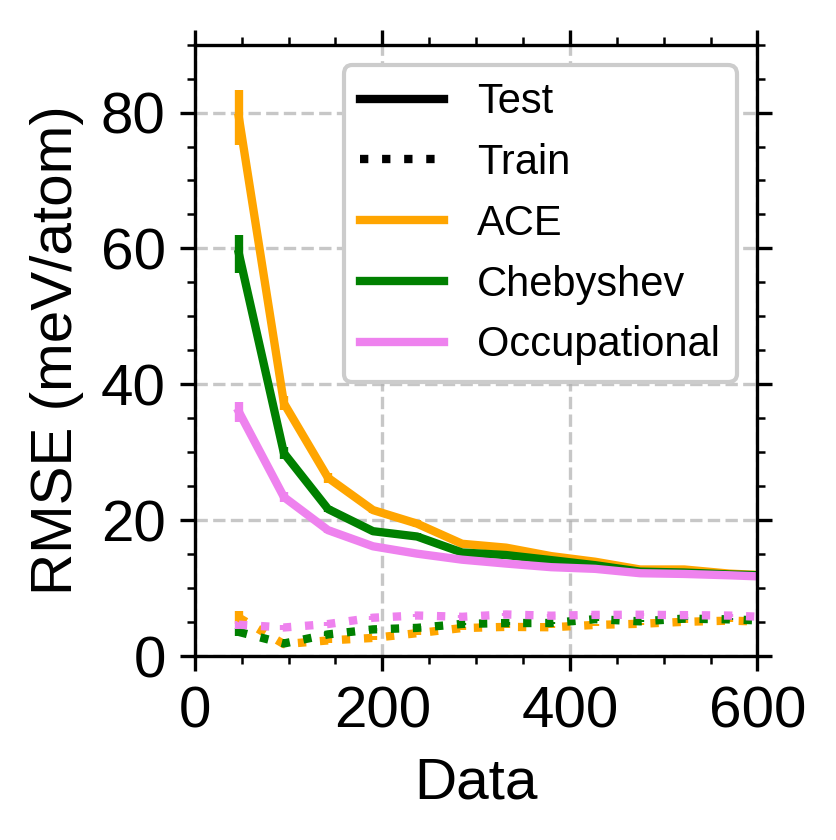}}\\
    \subfloat[][\label{fig:learning_curve_energy_EF}]{\includegraphics[height=0.5\linewidth]{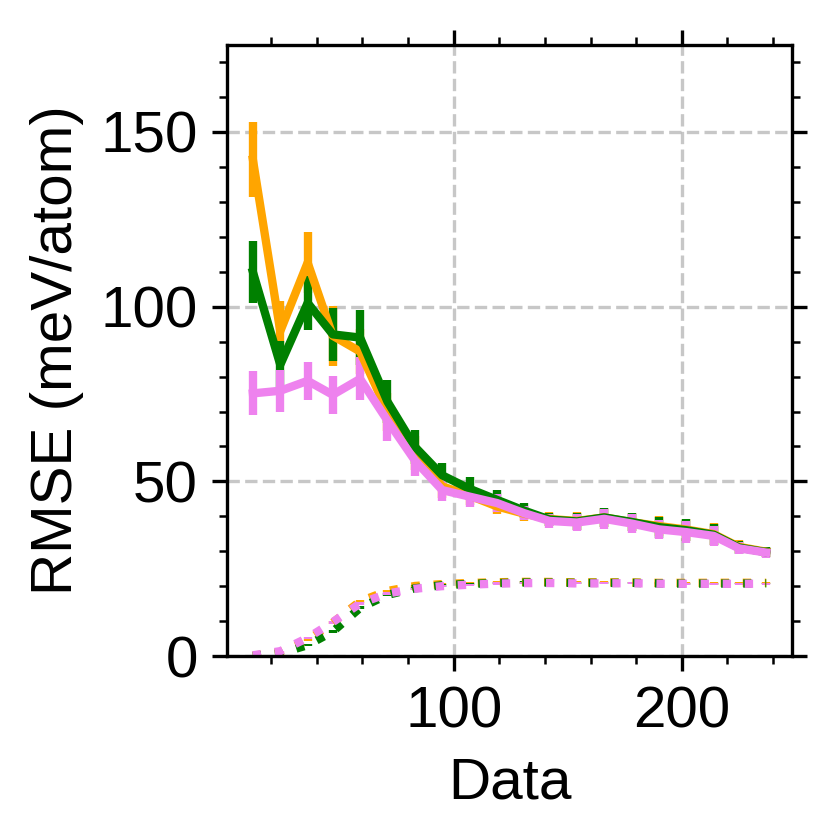}}
    \subfloat[][\label{fig:learning_curve_force_EF}]{\includegraphics[height=0.5\linewidth]{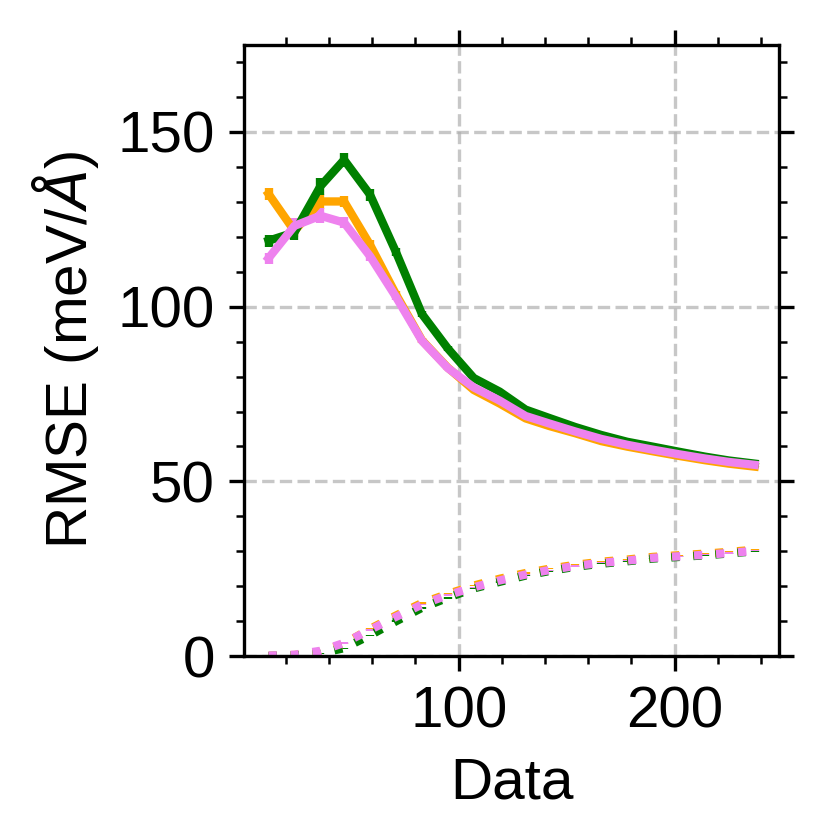}}
    \caption{(a) Convergence of test errors with the number of basis functions for fits to 80\% of the Mg--Nd dataset. Models to the left of the first dashed line contain only pair descriptors, those between the first and second dashed lines contain pair and triplet descriptors, and those beyond the second dashed line add quadruplet clusters to the most extensive triplet basis set. The curves are slightly shifted relative to each other for ease of visualization. (b--d) Learning curves of train and test errors for the quadruplet potential: (b) energy error for a fit to energies only, and (c) energy and (d) force errors for a fit to both energies and forces.}
    \label{fig:error_curves}
\end{figure}

\begin{figure*}[!ht]
    \centering
    \subfloat[][Pyramidal I \label{fig:Nd_pyramid1}]{\includegraphics[width=0.5\linewidth]{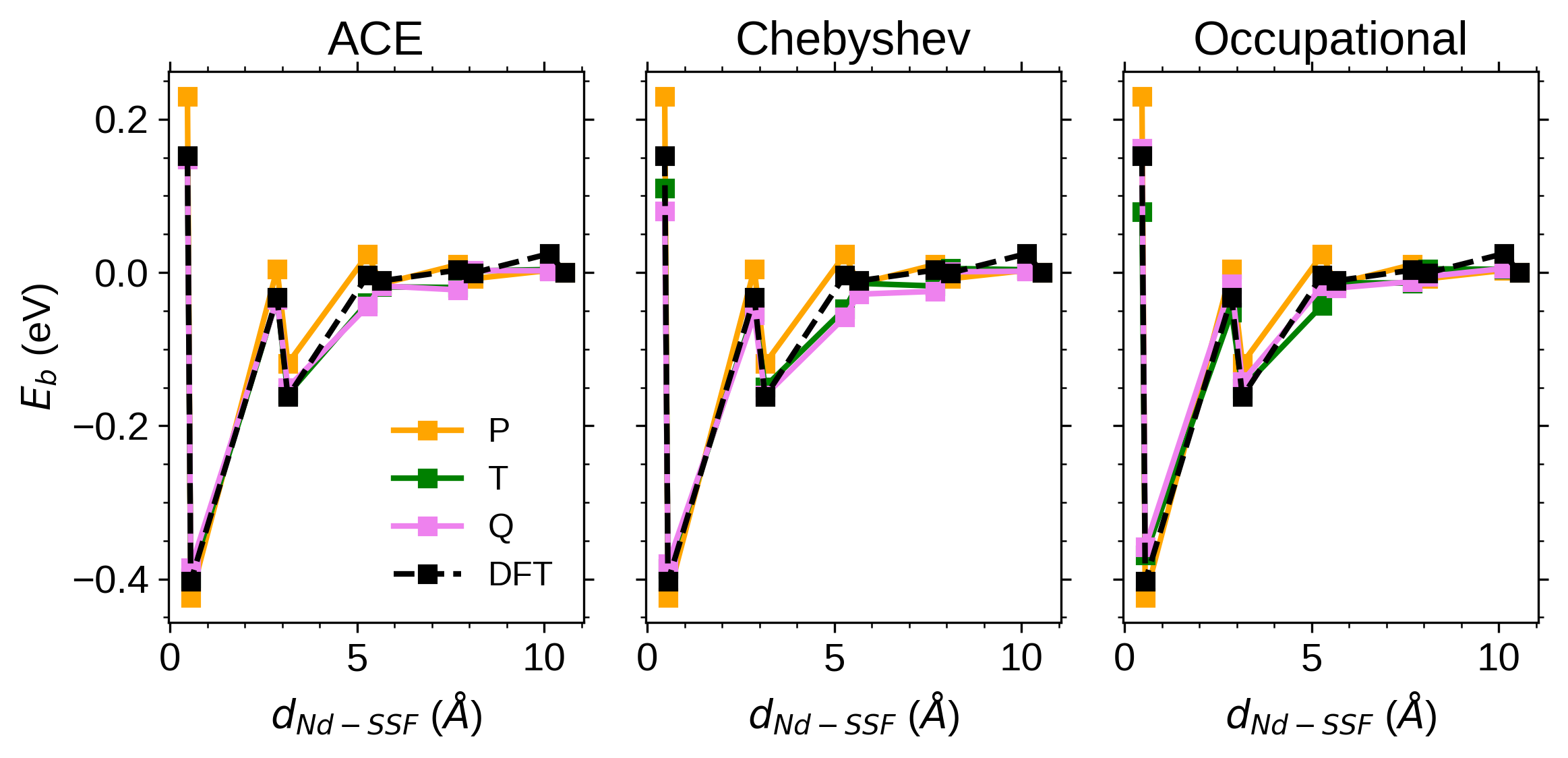}}
    \subfloat[][Basal \label{fig:Nd_basal}]{\includegraphics[width=0.5\linewidth]{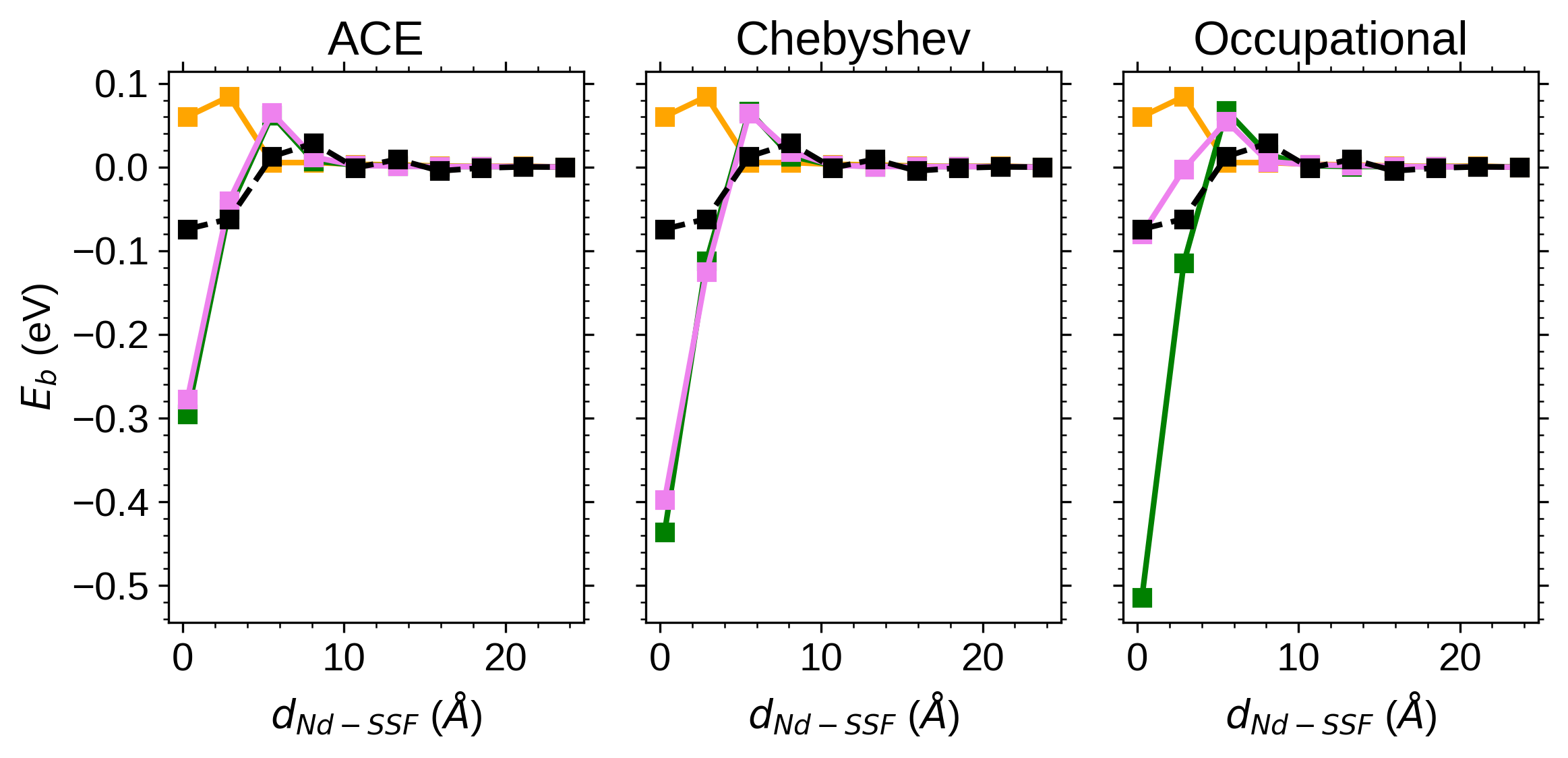}}
    \caption{Binding energy of a Nd solute to a pyramidal I (a) and a basal (b) stable stacking fault in HCP Mg, as a function of the distance between the solute and the fault. Within each panel, the three columns correspond to the conventional ACE, Chebyshev, and occupational site bases, and P, T, and Q denote the pair, triplet, and quadruplet potentials compared against the DFT reference. Negative energies indicate binding.}
\end{figure*}

\begin{figure}[!ht]
    \centering
    \includegraphics[width=0.8\linewidth]{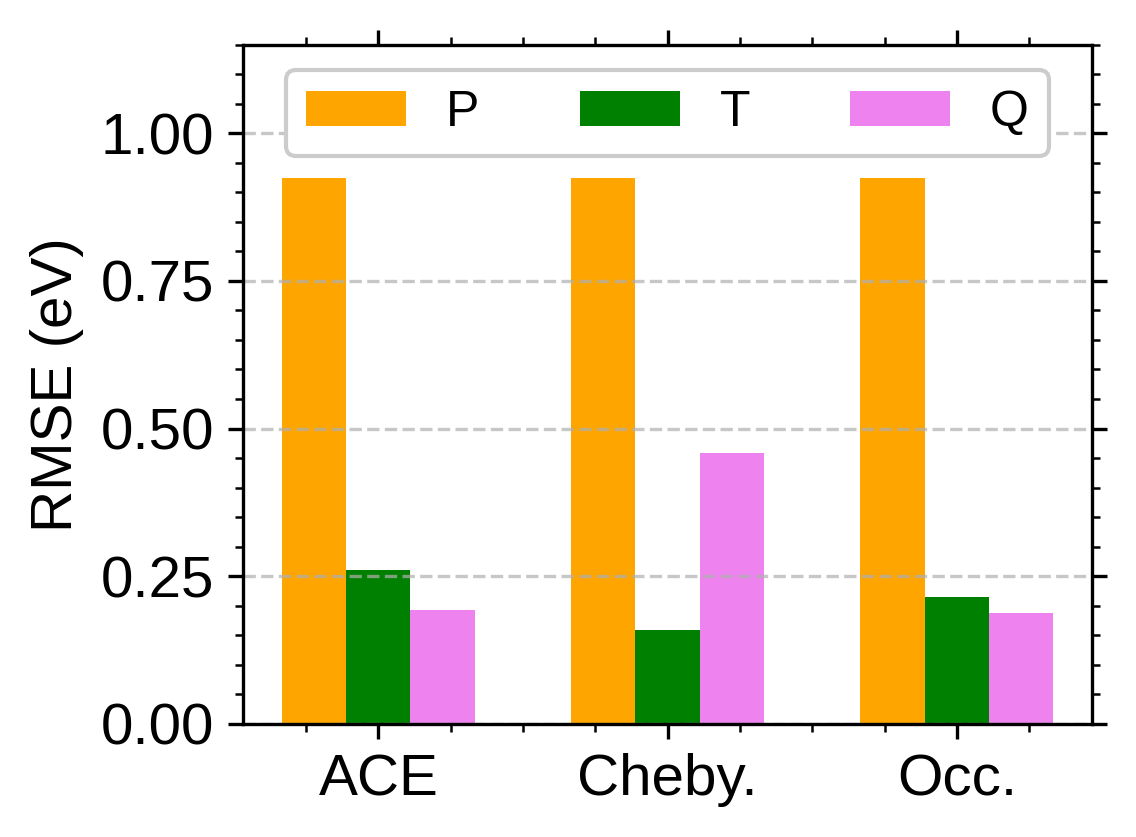}
    \caption{Root mean squared error in the predicted vacancy formation energies of several Mg--Nd phases, for pair (P), triplet (T), and quadruplet (Q) potentials built on the conventional ACE, Chebyshev, and occupational site bases.}
    \label{fig:vacancy_formation_energy_errors}
\end{figure}

\subsection{Dilute alloys} \label{sect:error_metrics_point_defects}
We begin with the influence of the chemical site basis on the defect properties of dilute alloys. Our case study is Mg--Nd, a magnesium alloy of interest for lightweight applications \cite{natarajan_early_2016,natarajan_unified_2017}, and we adopt a dataset from a previous study on this system \cite{piersante_machine_2026}. The reference dataset comprises 4746 Mg--Nd structures, primarily symmetry-distinct orderings on the HCP, BCC, and C15 Laves phase lattices. More than 90\% of these structures have a Nd composition below 50 at.\%. We examined the convergence of the root mean squared error (RMSE) with respect to the number of cluster basis functions and the size of the training set, averaging over 50 random train-test splits for each point in \cref{fig:error_curves}.

\Cref{fig:basis_functions} shows the convergence of the test errors as a function of the number of basis functions. The models were fitted to the energies of 80\% of the dataset, and the test RMSE was evaluated on the remaining 20\%. The curves begin with a pair potential and track the test error as triplet and then quadruplet cluster functions are added. Because the cluster basis is hierarchical, we refer to models truncated at pair, triplet, and quadruplet clusters as pair, triplet, and quadruplet potentials, labeled P, T, and Q in the figures. All three chemical bases follow similar trends, with the test error decreasing steadily and varying little with the choice of basis.

To explore the data-poor regime, we generated learning curves for the quadruplet potential with the largest basis set in \cref{fig:basis_functions}. In each case the test error was evaluated on the data left out of the training set. \Cref{fig:learning_curve_E} shows the learning curve for a model fitted to energies only. The occupational basis converges fastest, followed by the Chebyshev basis, and all three site bases reach the same test error at approximately 500 datapoints. \Cref{fig:learning_curve_energy_EF,fig:learning_curve_force_EF} show analogous learning curves for models trained on both energies and forces. The energy test error converges at a similar rate across the three bases, although the occupational basis remains more stable in the transient regime. The force test errors of the occupational and ACE bases are comparable, whereas the Chebyshev basis performs slightly worse. After the initial transient, all three bases attain the same energy and force test errors at around 120 datapoints. The reduced data requirement reflects the additional information supplied by the forces.

Statistical errors capture general trends, but the energy scale of many material properties is comparable to the RMSE, so validation against specific properties remains necessary \cite{piersante_machine_2026}. We next examined how the choice of chemical basis affects the prediction of defect properties in Mg--Nd alloys, focusing on the binding energy of a Nd solute to stable stacking faults (SSFs) in HCP Mg and on vacancy formation energies in various Mg--Nd phases \cite{delfino_phase_1990, natarajan_early_2016}. The training set for these benchmarks contains 145 crystal structures: HCP Mg supercells with an isolated Nd atom or a Nd pair, various SSF geometries, vacancy structures in HCP and BCC Mg, and several Mg--Nd phases. The dataset does not contain the DFT structures used to evaluate the defect properties below, so each model is tested on environments absent from its training set. We trained three models of increasing body order, from a pair potential to a quadruplet potential, fitting to both total energies and forces.

The binding energy of a solute atom to SSFs largely controls dislocation motion, including cross-slip \cite{wu_mechanistic_2018}. \Cref{fig:Nd_pyramid1,fig:Nd_basal} show the binding of a Nd atom to two SSFs in Mg as a function of distance. The three chemical bases give comparable results for the pyramidal I SSF. However, for the basal SSF, only the quadruplet potential with the occupational basis captures both the magnitude and the shape of the interaction. The ACE and Chebyshev bases generally predict excessively attractive binding energies at the stacking fault.

Vacancy properties similarly control diffusion in these alloys \cite{saal_solutevacancy_2012}. In particular, the difference in vacancy formation energy between matrix and precipitate phases influences the creep behavior of precipitation-hardened alloys \cite{choudhuri_exceptional_2017}. \Cref{fig:vacancy_formation_energy_errors} summarizes the RMSEs in the predicted vacancy formation energies across several Mg--Nd phases. The pair potentials give large errors for all three bases, as expected since pair potentials overestimate vacancy formation energies by roughly a factor of two \cite{carlsson_beyond_1990}. For the ACE and occupational bases, the RMSE decreases systematically as higher body orders are added. By contrast, the Chebyshev basis gives an RMSE that does not decrease monotonically with body order, and quadruplet potentials in this basis even predict negative vacancy formation energies, as shown in \cref{supp-fig:vacancy_parity_plots} of the Supporting Information.

\begin{figure*}[!ht]
    \centering
    \subfloat[][\label{fig:MoNb_mixing_enthalpy}]{\includegraphics[width=0.8\linewidth]{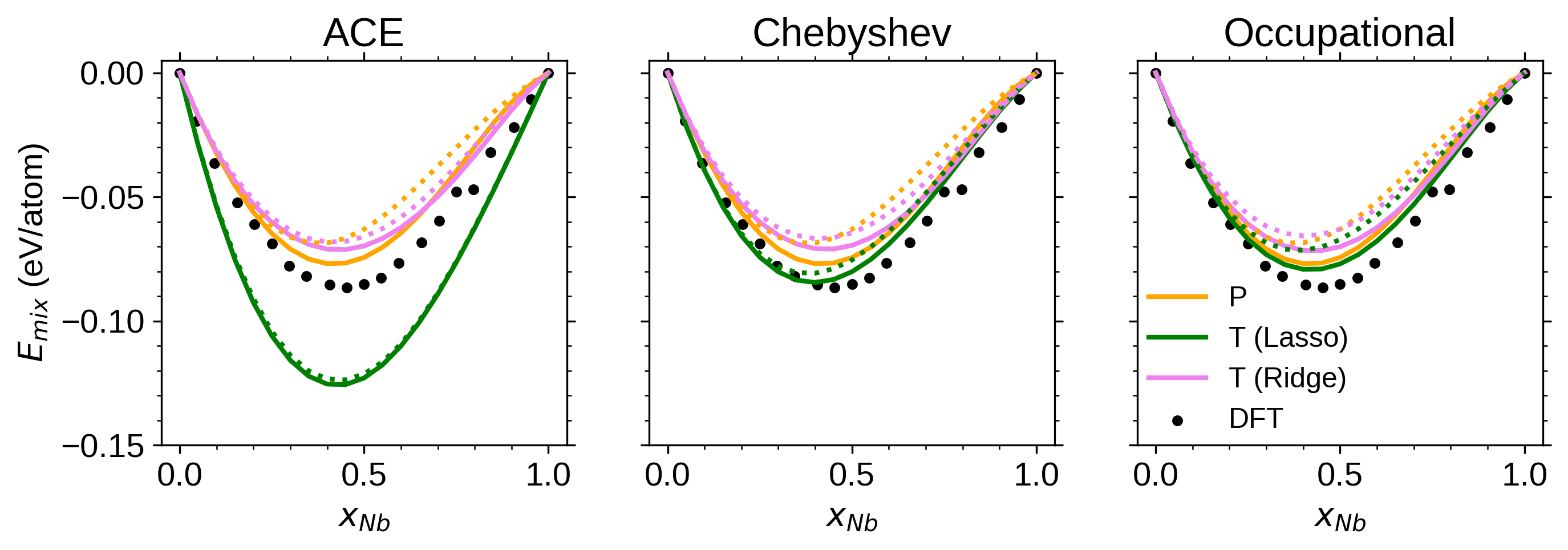}}\\
    \subfloat[][\label{fig:MoNb_mixing_enthalpy_convergence}]{\includegraphics[width=0.8\linewidth]{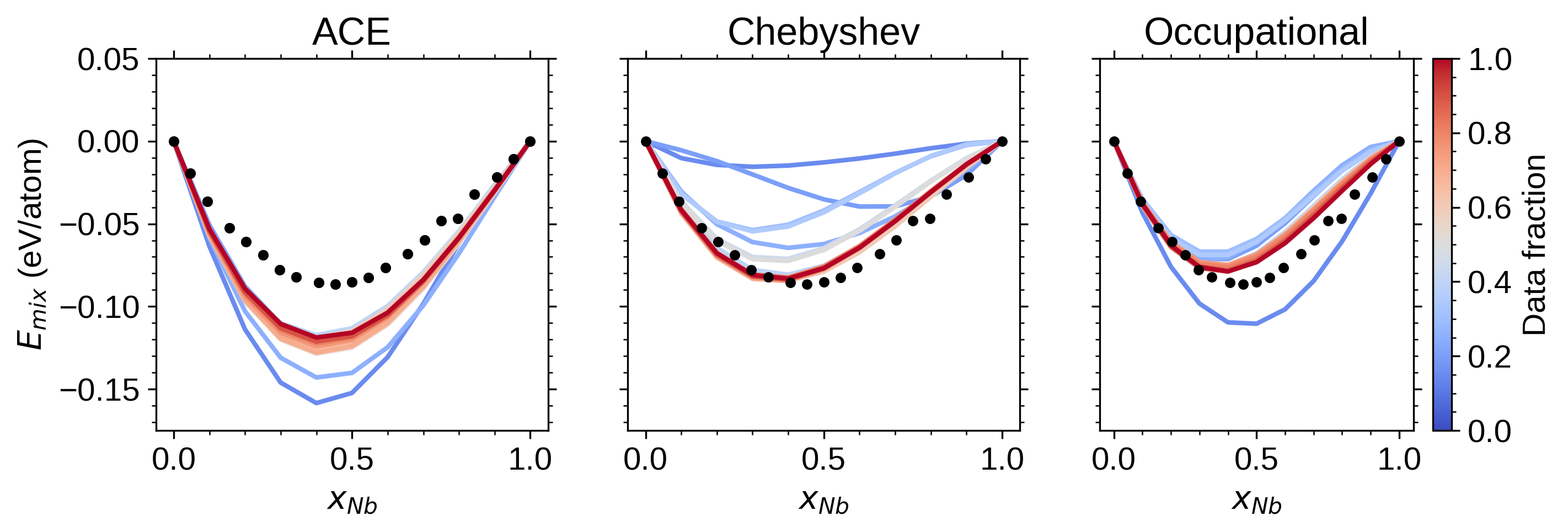}}
    \caption{Mo--Nb mixing properties. (a) Enthalpy predicted when fitting to all available data. Solid lines are predictions with the volume of the BCC cell relaxed to zero stress, and dotted lines are predictions with the volume following the rule of mixtures. Black dots are DFT estimates for special quasirandom structures, with volumes set by the rule of mixtures. (b) Convergence with the amount of training data of the enthalpy predicted by the T potential.}
\end{figure*}

\begin{figure*}[!ht]
    \centering
    \subfloat[][\label{fig:CrW_mixing_enthalpy}]{\includegraphics[width=0.8\linewidth]{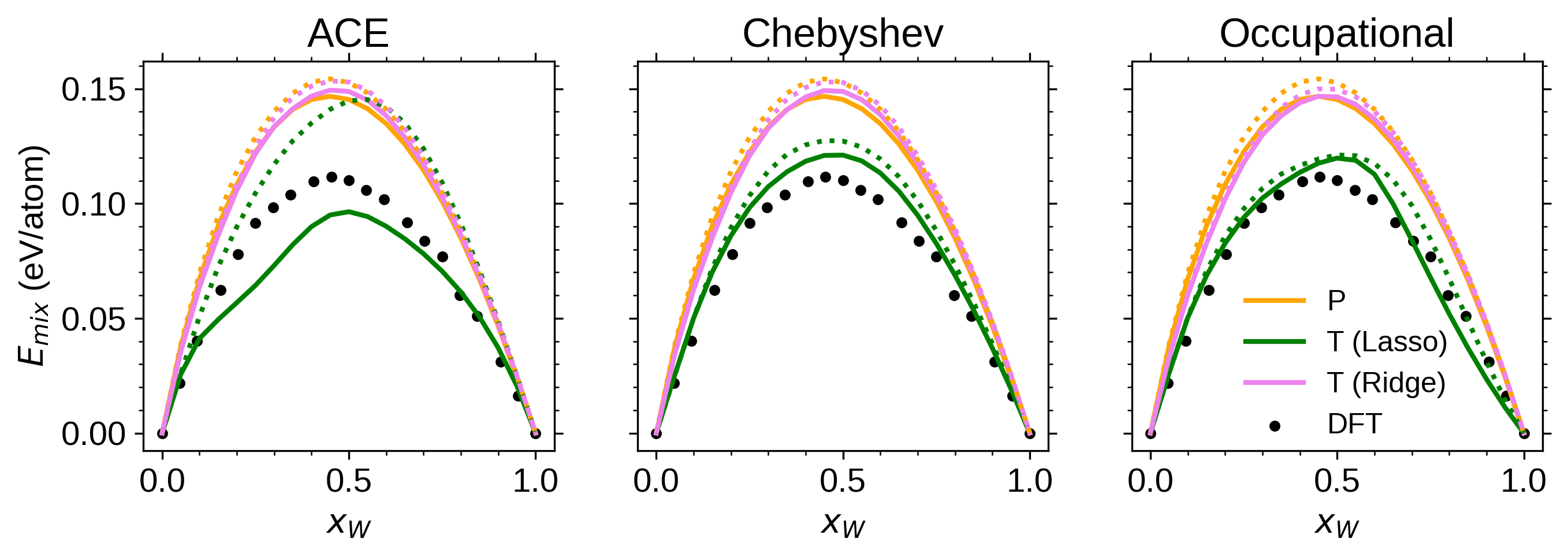}}\\
    \subfloat[][\label{fig:CrW_mixing_enthalpy_convergence}]{\includegraphics[width=0.8\linewidth]{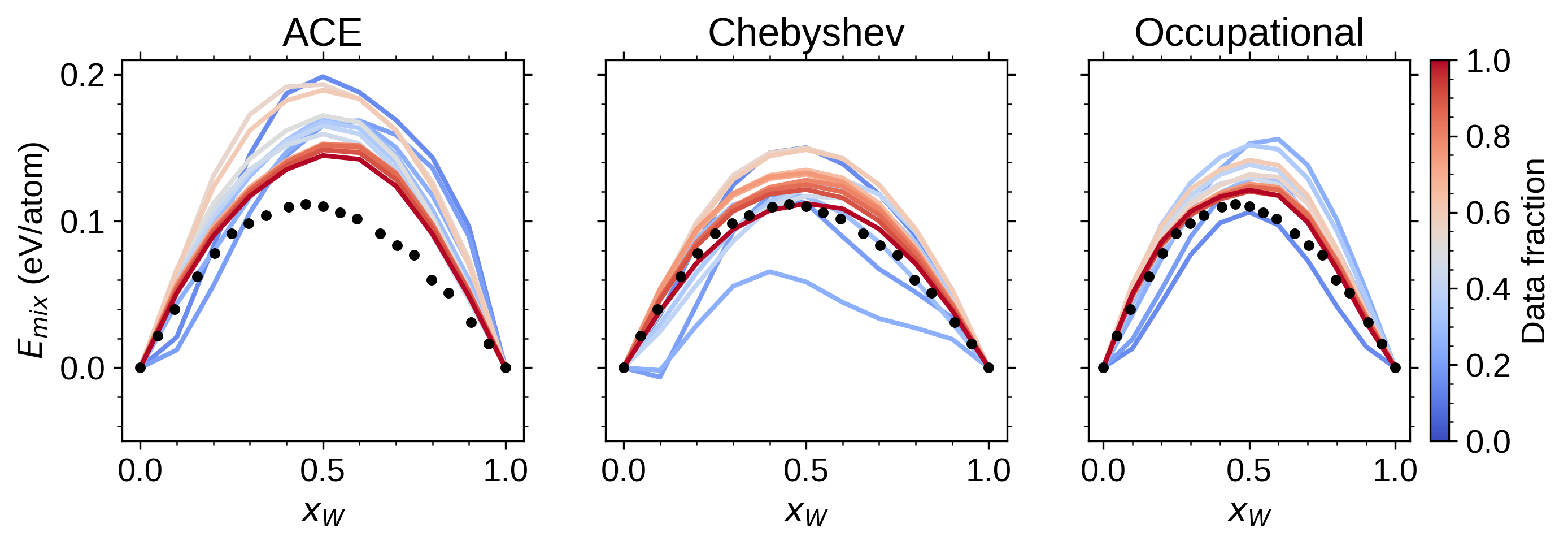}}
    \caption{Cr--W mixing properties. (a) Enthalpy predicted when fitting to all available data. Solid lines are predictions with the volume of the BCC cell relaxed to zero stress, and dotted lines are predictions with the volume following the rule of mixtures. Black dots are DFT estimates for special quasirandom structures, with volumes set by the rule of mixtures. (b) Convergence with the amount of training data of the enthalpy predicted by the T potential.}
\end{figure*}

\subsection{Concentrated random alloys} \label{sect:random_alloy}
The point-defect results show that capturing the interaction between a solute atom and a defect in a dilute alloy can require careful choice of the basis set size and the chemical basis. We next turn to the thermodynamic properties of concentrated disordered binary alloys. We investigated two prototypical refractory alloys, Mo--Nb and Cr--W, which are of interest for their high-temperature properties \cite{natarajan_crystallography_2020, lee_modeling_2026}. Mo--Nb forms ordered phases, indicating favorable interactions between Mo and Nb, whereas Cr--W exhibits a miscibility gap. The reference training sets for Mo--Nb and Cr--W contain 157 and 159 datapoints, respectively. The data comprise approximately 140 symmetry-distinct orderings enumerated on the BCC lattice, together with volumetric distortions of the pure elements about their equilibrium lattice parameters at 0\,K. For simplicity, we fitted the potentials to energies only. We repeated each fit ten times with independently drawn random training sets to quantify the variability of the fitted models. Each training set retained the volumetric distortions, so that the energy--volume dependence remained constrained in every fit. All figures report the mean prediction over the ten fitted models unless the entire dataset was used for training.

\Cref{fig:MoNb_mixing_enthalpy} compares the Mo--Nb mixing enthalpy predicted by pair and triplet average-atom potentials derived from a multicomponent ACE trained on the full dataset. The rule-of-mixtures predictions use endpoint volumes taken from the corresponding potential. The DFT reference was computed with special quasirandom structures (SQS) \cite{zunger_special_1990,puchala_casm_2023}, whose volumes were also adjusted according to the rule of mixtures. We fitted the pair potentials by ordinary least squares and the triplet potentials with either Ridge regression or Lasso. At the pair level, all three bases yield identical results. At the triplet level, Ridge regression produces very similar models across the three bases, all of which underestimate the mixing enthalpy. Lasso improves the predictions of the Chebyshev and occupational bases, but degrades the ACE basis model, which then strongly overestimates the mixing enthalpy. Overall, the triplet potential with the Chebyshev basis agrees most closely with the SQS data.

We next varied the size of the training set to test how quickly each basis converges. \Cref{fig:MoNb_mixing_enthalpy_convergence} tracks the mixing enthalpy as the number of training datapoints increases. For simplicity, the BCC cell volume was fixed according to the rule of mixtures using DFT-predicted endpoint volumes. The potentials were fitted using Lasso because it gave more accurate mixing enthalpies in \cref{fig:MoNb_mixing_enthalpy}. The ACE basis quickly converges to a mixing enthalpy that is too negative. The Chebyshev basis initially predicts a near-zero mixing enthalpy but rapidly converges towards the SQS estimate. The occupational basis exhibits the most stable convergence and reaches the SQS estimate with substantially fewer training datapoints than the other two bases.

\Cref{fig:CrW_mixing_enthalpy} shows the Cr--W mixing enthalpy predicted by pair and triplet average-atom potentials trained on the full dataset. In contrast to Mo--Nb, this system presents a miscibility gap. The 40\% volumetric misfit between BCC Cr and BCC W also makes the energy--volume dependence more difficult to learn. The three bases yield identical results at the pair level. The triplet potentials fitted with Ridge regression differ little across chemical bases and consistently overestimate the mixing enthalpy. With Lasso, the occupational and Chebyshev average-atom models reproduce the mixing enthalpy accurately. In contrast, the large separation between the zero-stress and rule-of-mixtures curves shows that the ACE basis does not capture the composition dependence of the molar volume, which the other two bases reproduce. \Cref{fig:CrW_mixing_enthalpy_convergence} shows the variation in Cr--W mixing enthalpy as the number of training datapoints increases for models fitted with Lasso. The ACE basis rapidly converges to a larger mixing enthalpy than estimated with SQS structures, whereas the occupational and Chebyshev bases converge to the SQS estimate. The occupational basis again shows the least variability across fits.

The better performance of the occupational and Chebyshev bases in \cref{fig:MoNb_mixing_enthalpy_convergence,fig:CrW_mixing_enthalpy_convergence} partly reflects their faster convergence with training set size. As shown in \cref{supp-fig:learning_curves_monb_crw_lasso} of the Supporting Information, the occupational basis achieves lower test errors for the smallest datasets, whereas both hierarchical bases attain similar errors as soon as the training set contains approximately 60 structures. By contrast, the ACE basis converges to higher RMSEs in both Mo--Nb and Cr--W. The variability of the fitted models also decreases as the training dataset grows, as reported in \cref{supp-fig:variance_convergence_lasso} of the Supporting Information. At equiatomic composition, and for potentials trained on 80--90\% of the data, the standard deviation of the Cr--W mixing enthalpy is 2--5\,meV/atom across the different chemical bases. In Mo--Nb, the standard deviation is 0.5--1\,meV/atom for the hierarchical bases, whereas it remains close to 2.5\,meV/atom for the ACE basis. These standard deviations are small compared with the magnitude of the mixing enthalpy itself, so the three bases agree once the training set is large. The differences between bases matter instead in the data-poor regime, where a careful choice of chemical basis delivers accurate disordered-alloy properties from smaller datasets and smaller basis sets.

\section{Discussion} \label{sect:discussion}
This study revisits the original formulation of the ACE \cite{drautz_atomic_2019,drautz_atomic_2020} to incorporate an arbitrary chemical site basis into the definition of the cluster functions. Accounting explicitly for the chemical degrees of freedom leads to the average-atom ACE for chemically disordered materials. Starting from the multicomponent ACE, we derive the average-atom site energy. The $B$-basis representation factorizes the mean-field energy into conventional and self-interaction contributions, with analytical expressions up to quadruplet clusters.

We benchmarked these models on a prototypical dilute Mg--Nd alloy and on two binary alloys that form disordered solid solutions at elevated temperatures. Reproducing the defect properties of the dilute alloy to \textit{ab initio} accuracy required care in the choice of both the chemical basis and the descriptor set. The same conclusion follows from the mixing enthalpies of the disordered solutions, evaluated with the average-atom potential of \cref{sect:average_atom_energy} from an ACE parameterized on a small dataset. Accurate disordered-alloy properties are therefore accessible from small datasets when the chemical basis and the descriptors are chosen carefully.

All three site bases reach comparable test errors when trained on 80\% of the Mg--Nd dataset, and distinct trends emerge only in the data-poor regime. The learning curves in \cref{sect:error_metrics_point_defects} show that the occupational basis converges fastest and most stably, resolving small energy differences even for small training sets. It also gives the most accurate point-defect binding energies and the most stable convergence of the mixing enthalpy in both Mo--Nb and Cr--W. The ACE and Chebyshev bases converge at comparable rates, and both struggle to reproduce the Nd binding energies to SSFs. Beyond that, the two bases fail on opposite tasks. The Chebyshev basis predicts the mixing enthalpies of the concentrated alloys well, but breaks down for the vacancy formation energies of the dilute alloy as the body order increases. The ACE basis captures the vacancy thermodynamics but misses the mixing enthalpies of Mo--Nb and Cr--W. The appropriate chemical basis therefore depends on the target property.

The origin of these differences remains unclear. Common practice in on-lattice cluster expansions favors the occupational basis for dilute alloys and the Chebyshev basis for concentrated alloys \cite{van_der_ven_nondilute_2008}. Our results follow this pattern in part, as the Chebyshev basis performs well for the concentrated alloys and poorly for several dilute-alloy properties. The difficulties of the ACE basis may reflect its overcompleteness. The ACE site basis effectively carries a third chemical component for which no training data exist, so the expansion covers only a subregion of the chemical space that its basis spans. This restriction may underlie its failure to reproduce the mixing enthalpies in \cref{fig:MoNb_mixing_enthalpy,fig:CrW_mixing_enthalpy}. The regularization may be another source of the discrepancy. Lasso was necessary to obtain accurate mixing enthalpies with the hierarchical bases in both Cr--W and Mo--Nb, which suggests that sparser weights suit the random alloy. The ACE basis, by contrast, did not reproduce the mixing enthalpy under either regularization. This may instead indicate that the ACE basis requires more complex descriptors to reproduce the disordered-phase interactions.

This work also illustrates that bespoke ACE models give direct access to thermodynamic quantities through their parameters. As shown in \cref{sect:average_atom,sect:random_alloy}, the mixing enthalpy of a solid solution can be extracted from a linear ACE model. Accurate estimates therefore require only a dataset of relatively small ordered structures rather than DFT calculations on large SQS cells.

\Cref{sect:self_interaction} derives closed-form expressions for the self-interactions of triplet and quadruplet clusters, beyond which the Clebsch-Gordan algebra becomes considerably more involved. Five-body self-interactions, for example, do not reduce to a lower-body-order $B$-basis function multiplied by an $l$-dependent prefactor. They require linear combinations of $B$-basis functions coupled through Wigner-$6j$ symbols, although each self-interaction term remains rotationally invariant on its own. Implementing these invariant features is therefore more cumbersome, but the general mathematical framework remains intact. The coincidence patterns created from $K$ bonds retain the structure of set partitions, and each pattern contributes one term to the unsymmetrized disordered-phase descriptor.

The number of terms in each average-atom descriptor follows the Bell numbers $\mathcal{B}_K$, which count the partitions of a set of $K$ elements. For $K$ from 2 to 5, the Bell numbers are 2, 5, 15, and 52. A cluster with $K$ bonds, and therefore body order $K+1$, contributes that many distinct terms to the unsymmetrized disordered-phase descriptor. Each term pairs a geometric descriptor with a chemical factor, which is a joint moment of the site basis functions that share a bond. Each set of coincident bonds maps to an atomic density evaluated on a product basis of bond functions. Together, these observations constrain the combinations of chemical factors and self-interacting $B$-basis functions allowed in the symmetrized disordered-phase descriptor. We therefore expect the following general form:
\begin{equation}
    \langle B_{\boldsymbol{\eta} \boldsymbol{\nu}} \rangle = \langle \sigma_{\eta_0} \rangle \prod_{k = 1}^K \langle \sigma_{\eta_k} \rangle B_{\boldsymbol{\nu}} + \sum_{\boldsymbol{b} \in \mathcal{P}(\boldsymbol{\eta})} S_{\boldsymbol{b}}^{\text{chem.}} \sum_{\tilde{\boldsymbol{\nu}}(\boldsymbol{b})}C_{\tilde{\boldsymbol{\nu}}(\boldsymbol{b})}^{\text{ang.}} B_{\tilde{\boldsymbol{\nu}}(\boldsymbol{b})}^{\text{self}}
\end{equation}
where $\mathcal{P}(\boldsymbol{\eta})$ denotes the set partitions of $K$ bonds excluding the case of totally distinct bonds, $S_{\boldsymbol{b}}^{\text{chem.}}$ is the chemical factor associated with the coincidence pattern $\boldsymbol{b}$, and $C_{\tilde{\boldsymbol{\nu}}(\boldsymbol{b})}^{\text{ang.}}$ is the $l$-dependent factor attached to each permitted self-interacting $B$-basis function. The $B_{\boldsymbol{\nu}}$ are rotationally invariant descriptors built from the standard ACE $A$-basis. The $B_{\tilde{\boldsymbol{\nu}}}^{\text{self}}$ are likewise rotationally invariant, but they combine the standard $A$-basis with an augmented $A$-basis in which radial functions are repeated on the same bond.
The prefactors associated with the self-interactions could be determined through data-driven approaches. For instance, in a study on the conventional ACE, Ho \textit{et al.} \cite{ho_atomic_2024} used linear regression to eliminate the self-interactions from the ordinary ACE descriptors. Establishing this general form beyond quadruplet clusters, and determining the angular prefactors that enter it, remains a challenge for future work.

\section{Conclusions}
In this article, we generalize the multicomponent ACE to arbitrary chemical site basis functions. From there, we derive the average-atom ACE for random alloys as the composition-weighted mean-field approximation of the multicomponent ACE. This procedure produces an ACE with composition-dependent coefficients, augmented with descriptors that encode the self-interactions between bonds. Our exact treatment of the self-interactions preserves the linear scaling of the ACE, which is necessary for an efficient implementation of average-atom potentials. The mapping between average-atom coefficients and the coefficients of a fitted ACE also allows any fitted model to be converted into its average-atom counterpart. Examples ranging from dilute point defects to concentrated alloys demonstrate that the chemical site basis is a design choice that governs data efficiency. The occupational basis performs best in the data-poor regime, while the three bases give comparable results once data are plentiful. In future work, we will apply the average-atom framework to baseline properties of realistic compositionally complex alloys at \textit{ab initio} accuracy.

\section{Associated Content}
\subsection{Supporting information}
The Supporting Information is available free of charge at url-to-be-inserted-by-the-publisher.

Derivations of disordered quadruplet descriptor, derivation of $l$-dependent prefactors for triplet and quadruplet self-interactions, DFT settings, training and validation datasets, additional results on test error convergence, point defects, disordered alloy properties (PDF).

\subsection{Data Availability Statement}
The data associated with this study are available at \cite{data_avail}.

\section{Acknowledgments}
This research was supported by the Swiss National Science Foundation (Grant No. 215178). The authors are very grateful to Joseph Abbott, Davide Tisi, and Guillaume Fraux for their help with \verb|featomic| and \verb|metatensor|. The authors acknowledge the use of \verb|pymatgen| \cite{ong_python_2013} and \verb|ase| \cite{hjorth_larsen_atomic_2017}.

\bibliographystyle{unsrt}
\bibliography{bibliography}

\begin{thebibliography}{10}

\bibitem{deringer_machine_2019}
Volker~L. Deringer, Miguel~A. Caro, and Gábor Csányi.
\newblock Machine {Learning} {Interatomic} {Potentials} as {Emerging} {Tools}
  for {Materials} {Science}.
\newblock {\em Advanced Materials}, 31(46):1902765, November 2019.

\bibitem{friederich_machine-learned_2021}
Pascal Friederich, Florian Häse, Jonny Proppe, and Alán Aspuru-Guzik.
\newblock Machine-learned potentials for next-generation matter simulations.
\newblock {\em Nature Materials}, 20(6):750--761, June 2021.

\bibitem{guo_intercalation_2023}
Xingyu Guo, Chi Chen, and Shyue~Ping Ong.
\newblock Intercalation {Chemistry} of the {Disordered} {Rocksalt}
  {Li}$_{\textrm{3}}$ {V}$_{\textrm{2}}$ {O}$_{\textrm{5}}$ {Anode} from
  {Cluster} {Expansions} and {Machine} {Learning} {Interatomic} {Potentials}.
\newblock {\em Chemistry of Materials}, 35(4):1537--1546, February 2023.

\bibitem{deringer_modelling_2020}
Volker~L Deringer.
\newblock Modelling and understanding battery materials with
  machine-learning-driven atomistic simulations.
\newblock {\em Journal of Physics: Energy}, 2(4):041003, October 2020.

\bibitem{eyert_machine-learned_2023}
Volker Eyert, Jonathan Wormald, William~A. Curtin, and Erich Wimmer.
\newblock Machine-learned interatomic potentials: {Recent} developments and
  prospective applications.
\newblock {\em Journal of Materials Research}, 38(24):5079--5094, December
  2023.

\bibitem{rahman_review_2025}
Arafat Rahman, Md~Sojib Hossain, and Abdullah-Bin Siddique.
\newblock Review: machine learning approaches for diverse alloy systems.
\newblock {\em Journal of Materials Science}, 60(29):12189--12221, August 2025.

\bibitem{mazitov_pet-mad_2025}
Arslan Mazitov, Filippo Bigi, Matthias Kellner, Paolo Pegolo, Davide Tisi,
  Guillaume Fraux, Sergey Pozdnyakov, Philip Loche, and Michele Ceriotti.
\newblock {PET}-{MAD} as a lightweight universal interatomic potential for
  advanced materials modeling.
\newblock {\em Nature Communications}, 16(1):10653, November 2025.

\bibitem{marchand_foundation_2025}
Daniel Marchand.
\newblock Foundation models for metallurgy?
\newblock {\em MRS Bulletin}, 50(7):805--818, July 2025.

\bibitem{lysogorskiy_graph_2026}
Yury Lysogorskiy, Anton Bochkarev, and Ralf Drautz.
\newblock Graph atomic cluster expansion for foundational machine learning
  interatomic potentials.
\newblock {\em npj Computational Materials}, February 2026.

\bibitem{radova_fine-tuning_2025}
Mariia Radova, Wojciech~G. Stark, Connor~S. Allen, Reinhard~J. Maurer, and
  Albert~P. Bartók.
\newblock Fine-tuning foundation models of materials interatomic potentials
  with frozen transfer learning.
\newblock {\em npj Computational Materials}, 11(1):237, July 2025.

\bibitem{wong_bias_2026}
Nicolas~H. Wong and Julia~H. Yang.
\newblock Bias in universal machine-learned interatomic potentials and its
  effects on fine-tuning.
\newblock {\em Journal of Chemical Theory and Computation}, 22(13):6820--6834,
  06 2026.

\bibitem{toit_hyperparameter_2024}
Daniel~F. Thomas~du Toit, Yuxing Zhou, and Volker~L. Deringer.
\newblock Hyperparameter optimization for atomic cluster expansion potentials.
\newblock {\em Journal of Chemical Theory and Computation},
  20(22):10103--10113, 11 2024.

\bibitem{piersante_machine_2026}
Lorenzo Piersante and Anirudh~Raju Natarajan.
\newblock Machine learning interatomic potentials for solid-state
  precipitation.
\newblock {\em Phys. Rev. Mater.}, 10:093802, Sep 2026.

\bibitem{drautz_atomic_2019}
Ralf Drautz.
\newblock Atomic cluster expansion for accurate and transferable interatomic
  potentials.
\newblock {\em Physical Review B}, 99(1):014104, January 2019.

\bibitem{drautz_atomic_2020}
Ralf Drautz.
\newblock Atomic cluster expansion of scalar, vectorial, and tensorial
  properties including magnetism and charge transfer.
\newblock {\em Physical Review B}, 102(2):024104, July 2020.

\bibitem{lysogorskiy_performant_2021}
Yury Lysogorskiy, Cas Van~Der Oord, Anton Bochkarev, Sarath Menon, Matteo
  Rinaldi, Thomas Hammerschmidt, Matous Mrovec, Aidan Thompson, Gábor Csányi,
  Christoph Ortner, and Ralf Drautz.
\newblock Performant implementation of the atomic cluster expansion ({PACE})
  and application to copper and silicon.
\newblock {\em npj Computational Materials}, 7(1):97, June 2021.

\bibitem{bochkarev_efficient_2022}
Anton Bochkarev, Yury Lysogorskiy, Sarath Menon, Minaam Qamar, Matous Mrovec,
  and Ralf Drautz.
\newblock Efficient parametrization of the atomic cluster expansion.
\newblock {\em Physical Review Materials}, 6(1):013804, January 2022.

\bibitem{ibrahim_atomic_2023}
Eslam Ibrahim, Yury Lysogorskiy, Matous Mrovec, and Ralf Drautz.
\newblock Atomic cluster expansion for a general-purpose interatomic potential
  of magnesium.
\newblock {\em Physical Review Materials}, 7(11):113801, November 2023.

\bibitem{qamar_atomic_2023}
Minaam Qamar, Matous Mrovec, Yury Lysogorskiy, Anton Bochkarev, and Ralf
  Drautz.
\newblock Atomic cluster expansion for quantum-accurate large-scale simulations
  of carbon.
\newblock {\em Journal of Chemical Theory and Computation}, 19(15):5151--5167,
  06 2023.

\bibitem{ibrahim_atomic_2026}
Eslam Ibrahim, Yury Lysogorskiy, Ralf Drautz, and Pablo~M. Piaggi.
\newblock Water phase diagram from a general-purpose atomic cluster expansion
  potential.
\newblock {\em Journal of Chemical Theory and Computation}, 22(9):4758--4766,
  04 2026.

\bibitem{bienvenu_development_2025}
Baptiste Bienvenu, Mira Todorova, Jörg Neugebauer, Dierk Raabe, Matous Mrovec,
  Yury Lysogorskiy, and Ralf Drautz.
\newblock Development of an atomic cluster expansion potential for iron and its
  oxides.
\newblock {\em npj Computational Materials}, 11(1):81, March 2025.

\bibitem{sanchez_generalized_1984}
J.M. Sanchez, F.~Ducastelle, and D.~Gratias.
\newblock Generalized cluster description of multicomponent systems.
\newblock {\em Physica A: Statistical Mechanics and its Applications},
  128(1-2):334--350, November 1984.

\bibitem{sanchez_cluster_2010}
J.~M. Sanchez.
\newblock Cluster expansion and the configurational theory of alloys.
\newblock {\em Physical Review B}, 81(22):224202, June 2010.

\bibitem{dusson_atomic_2022}
Geneviève Dusson, Markus Bachmayr, Gábor Csányi, Ralf Drautz, Simon Etter,
  Cas Van Der~Oord, and Christoph Ortner.
\newblock Atomic cluster expansion: {Completeness}, efficiency and stability.
\newblock {\em Journal of Computational Physics}, 454:110946, April 2022.

\bibitem{smith_application_1989}
Richard~W. Smith and Gary~S. Was.
\newblock Application of molecular dynamics to the study of hydrogen
  embrittlement in {Ni}-{Cr}-{Fe} alloys.
\newblock {\em Physical Review B}, 40(15):10322--10336, November 1989.

\bibitem{varvenne_average-atom_2016}
Céline Varvenne, Aitor Luque, Wolfram~G. Nöhring, and William~A. Curtin.
\newblock Average-atom interatomic potential for random alloys.
\newblock {\em Physical Review B}, 93(10):104201, March 2016.

\bibitem{hodapp_exact_2025}
M.~Hodapp.
\newblock Exact average many-body interatomic interaction model for random
  alloys.
\newblock {\em Computational Materials Today}, 5:100018, March 2025.

\bibitem{zunger_special_1990}
Alex Zunger, S.-H. Wei, L.~G. Ferreira, and James~E. Bernard.
\newblock Special quasirandom structures.
\newblock {\em Physical Review Letters}, 65(3):353--356, July 1990.

\bibitem{goff_permutation-adapted_2024}
J.M. Goff, C.~Sievers, M.A. Wood, and A.P. Thompson.
\newblock Permutation-adapted complete and independent basis for atomic cluster
  expansion descriptors.
\newblock {\em Journal of Computational Physics}, 510:113073, August 2024.

\bibitem{barroso-luque_cluster_2024}
Luis Barroso-Luque and Gerbrand Ceder.
\newblock The cluster decomposition of the configurational energy of
  multicomponent alloys.
\newblock {\em npj Computational Materials}, 10(1):158, July 2024.

\bibitem{ho_atomic_2024}
Cheuk~Hin Ho, Timon~S. Gutleb, and Christoph Ortner.
\newblock Atomic cluster expansion without self-interaction.
\newblock {\em Journal of Computational Physics}, 515:113271, October 2024.

\bibitem{varshalovich_quantum_1988}
D.~A. Varshalovich, A.~N. Moskalev, and V.~K. Khersonskii.
\newblock {\em Quantum {Theory} of {Angular} {Momentum}}.
\newblock World Scientific Publishing, 1988.

\bibitem{edmonds_angular_1960}
A.~R. Edmonds.
\newblock {\em Angular {Momentum} in {Quantum} {Mechanics}}.
\newblock Princeton University Press, Princeton, New Jersey, second edition,
  1960.

\bibitem{kresse_ab_1993}
G.~Kresse and J.~Hafner.
\newblock \textit{{Ab} initio} molecular dynamics for liquid metals.
\newblock {\em Physical Review B}, 47(1):558--561, January 1993.

\bibitem{kresse_ab_1994}
G.~Kresse and J.~Hafner.
\newblock \textit{{Ab} initio} molecular-dynamics simulation of the
  liquid-metal–amorphous-semiconductor transition in germanium.
\newblock {\em Physical Review B}, 49(20):14251--14269, May 1994.

\bibitem{kresse_efficient_1996}
G.~Kresse and J.~Furthmüller.
\newblock Efficient iterative schemes for \textit{ab initio} total-energy
  calculations using a plane-wave basis set.
\newblock {\em Physical Review B}, 54(16):11169--11186, October 1996.

\bibitem{kresse_ultrasoft_1999}
G.~Kresse and D.~Joubert.
\newblock From ultrasoft pseudopotentials to the projector augmented-wave
  method.
\newblock {\em Physical Review B}, 59(3):1758--1775, January 1999.

\bibitem{perdew_generalized_1996}
John~P. Perdew, Kieron Burke, and Matthias Ernzerhof.
\newblock Generalized {Gradient} {Approximation} {Made} {Simple}.
\newblock {\em Physical Review Letters}, 77(18):3865--3868, October 1996.

\bibitem{perdew_generalized_1997}
John~P. Perdew, Kieron Burke, and Matthias Ernzerhof.
\newblock Generalized {Gradient} {Approximation} {Made} {Simple} [{Phys}.
  {Rev}. {Lett}. 77, 3865 (1996)].
\newblock {\em Physical Review Letters}, 78(7):1396--1396, February 1997.

\bibitem{lee_modeling_2026}
Damien~K.J. Lee, Yann~L. Müller, and Anirudh~Raju Natarajan.
\newblock Modeling the equilibrium vacancy concentration in multi-principal
  element alloys from first-principles.
\newblock {\em Acta Materialia}, 304:121752, January 2026.

\bibitem{bigi_fast_2023}
Filippo Bigi, Guillaume Fraux, Nicholas~J. Browning, and Michele Ceriotti.
\newblock Fast evaluation of spherical harmonics with sphericart.
\newblock {\em The Journal of Chemical Physics}, 159(6):064802, August 2023.

\bibitem{bigi_metatensor_2026}
Filippo Bigi, Joseph~W. Abbott, Philip Loche, Arslan Mazitov, Davide Tisi,
  Marcel~F. Langer, Alexander Goscinski, Paolo Pegolo, Sanggyu Chong, Rohit
  Goswami, Pol Febrer, Sofiia Chorna, Matthias Kellner, Michele Ceriotti, and
  Guillaume Fraux.
\newblock metatensor and metatomic : {Foundational} libraries for interoperable
  atomistic machine learning.
\newblock {\em The Journal of Chemical Physics}, 164(6):064113, February 2026.

\bibitem{natarajan_early_2016}
Anirudh~Raju Natarajan, Ellen~L.S. Solomon, Brian Puchala, Emmanuelle~A.
  Marquis, and Anton Van Der~Ven.
\newblock On the early stages of precipitation in dilute {Mg}–{Nd} alloys.
\newblock {\em Acta Materialia}, 108:367--379, April 2016.

\bibitem{natarajan_unified_2017}
Anirudh~Raju Natarajan and Anton Van Der~Ven.
\newblock A unified description of ordering in {HCP} {Mg}-{RE} alloys.
\newblock {\em Acta Materialia}, 124:620--632, February 2017.

\bibitem{delfino_phase_1990}
S.~Delfino, A.~Saccone, and R.~Ferro.
\newblock Phase relationships in the neodymium-magnesium alloy system.
\newblock {\em Metallurgical Transactions A}, 21(8):2109--2114, August 1990.

\bibitem{wu_mechanistic_2018}
Zhaoxuan Wu, Rasool Ahmad, Binglun Yin, Stefanie Sandlöbes, and W.~A. Curtin.
\newblock Mechanistic origin and prediction of enhanced ductility in magnesium
  alloys.
\newblock {\em Science}, 359(6374):447--452, January 2018.

\bibitem{saal_solutevacancy_2012}
James~E. Saal and C.~Wolverton.
\newblock Solute–vacancy binding of the rare earths in magnesium from first
  principles.
\newblock {\em Acta Materialia}, 60(13-14):5151--5159, August 2012.

\bibitem{choudhuri_exceptional_2017}
Deep Choudhuri, Srivilliputhur~G. Srinivasan, Mark~A. Gibson, Yufeng Zheng,
  David~L. Jaeger, Hamish~L. Fraser, and Rajarshi Banerjee.
\newblock Exceptional increase in the creep life of magnesium rare-earth alloys
  due to localized bond stiffening.
\newblock {\em Nature Communications}, 8(1):2000, December 2017.

\bibitem{carlsson_beyond_1990}
AE~Carlsson.
\newblock {\em Beyond {Pair} {Potentials} in {Elemental} {Transition} {Metals}
  and {Semiconductors}}, volume~43 of {\em Solid {State} {Physics}}.
\newblock Academic Press, New York London [etc.], 1990.
\newblock Publication Title: Solid state physics advances in research and
  applications.

\bibitem{natarajan_crystallography_2020}
Anirudh~Raju Natarajan, Pavel Dolin, and Anton Van Der~Ven.
\newblock Crystallography, thermodynamics and phase transitions in refractory
  binary alloys.
\newblock {\em Acta Materialia}, 200:171--186, November 2020.

\bibitem{puchala_casm_2023}
Brian Puchala, John~C. Thomas, Anirudh~Raju Natarajan, Jon~Gabriel Goiri,
  Sesha~Sai Behara, Jonas~L. Kaufman, and Anton Van Der~Ven.
\newblock {CASM} — {A} software package for first-principles based study of
  multicomponent crystalline solids.
\newblock {\em Computational Materials Science}, 217:111897, January 2023.

\bibitem{van_der_ven_nondilute_2008}
Anton Van Der~Ven, John~C. Thomas, Qingchuan Xu, Benjamin Swoboda, and Dane
  Morgan.
\newblock Nondilute diffusion from first principles: {Li} diffusion in {Li} x
  {TiS} 2.
\newblock {\em Physical Review B}, 78(10):104306, September 2008.

\bibitem{data_avail}
{L. Piersante} and {A. R. Natarajan}, 2026, {Materials Cloud},
  https://doi.org/10.24435/materialscloud:8m-rp.

\bibitem{ong_python_2013}
Shyue~Ping Ong, William~Davidson Richards, Anubhav Jain, Geoffroy Hautier,
  Michael Kocher, Shreyas Cholia, Dan Gunter, Vincent~L. Chevrier, Kristin~A.
  Persson, and Gerbrand Ceder.
\newblock Python {Materials} {Genomics} (pymatgen): {A} robust, open-source
  python library for materials analysis.
\newblock {\em Computational Materials Science}, 68:314--319, February 2013.

\bibitem{hjorth_larsen_atomic_2017}
Ask Hjorth~Larsen, Jens Jørgen~Mortensen, Jakob Blomqvist, Ivano~E Castelli,
  Rune Christensen, Marcin Dułak, Jesper Friis, Michael~N Groves, Bjørk
  Hammer, Cory Hargus, Eric~D Hermes, Paul~C Jennings, Peter Bjerre~Jensen,
  James Kermode, John~R Kitchin, Esben Leonhard~Kolsbjerg, Joseph Kubal,
  Kristen Kaasbjerg, Steen Lysgaard, Jón Bergmann~Maronsson, Tristan Maxson,
  Thomas Olsen, Lars Pastewka, Andrew Peterson, Carsten Rostgaard, Jakob
  Schiøtz, Ole Schütt, Mikkel Strange, Kristian~S Thygesen, Tejs Vegge, Lasse
  Vilhelmsen, Michael Walter, Zhenhua Zeng, and Karsten~W Jacobsen.
\newblock The atomic simulation environment—a {Python} library for working
  with atoms.
\newblock {\em Journal of Physics: Condensed Matter}, 29(27):273002, July 2017.

\end{thebibliography}


\begin{thebibliography}{1}

\bibitem{varshalovich_quantum_1988}
D.~A. Varshalovich, A.~N. Moskalev, and V.~K. Khersonskii.
\newblock {\em Quantum {Theory} of {Angular} {Momentum}}.
\newblock World Scientific Publishing, 1988.

\bibitem{edmonds_angular_1960}
A.~R. Edmonds.
\newblock {\em Angular {Momentum} in {Quantum} {Mechanics}}.
\newblock Princeton University Press, Princeton, New Jersey, second edition,
  1960.

\bibitem{piersante_machine_2026}
Lorenzo Piersante and Anirudh~Raju Natarajan.
\newblock Machine learning interatomic potentials for solid-state
  precipitation.
\newblock {\em Phys. Rev. Mater.}, 10:093802, Sep 2026.

\bibitem{lee_modeling_2026}
Damien~K.J. Lee, Yann~L. Müller, and Anirudh~Raju Natarajan.
\newblock Modeling the equilibrium vacancy concentration in multi-principal
  element alloys from first-principles.
\newblock {\em Acta Materialia}, 304:121752, January 2026.

\end{thebibliography}

\end{document}

\title{Supporting Information: Chemical site bases and average-atom potentials for the atomic cluster expansion}

\author{Lorenzo Piersante}
\affiliation{Laboratory of Materials Design and Simulation (MADES), Institute of Materials, \'{E}cole Polytechnique F\'{e}d\'{e}rale de Lausanne}

\author{Anirudh Raju Natarajan}
\email{anirudh.natarajan@epfl.ch}
\affiliation{Laboratory of Materials Design and Simulation (MADES), Institute of Materials, \'{E}cole Polytechnique F\'{e}d\'{e}rale de Lausanne}

\date{\today}

\maketitle

\section{Disordered-phase quadruplet descriptor} \label{sect:quadruplet_descriptors}
This section derives the average-atom descriptor of a generic quadruplet cluster. The $A$-basis function of a quadruplet is:
\begin{equation}
    A_{\boldsymbol{\eta}\boldsymbol{\nu}} = \sigma_{\eta_0}(\mu_0) A_{\eta_1 \nu_1} A_{\eta_2 \nu_2} A_{\eta_3 \nu_3}
\end{equation}
Averaging over the disordered chemical configurations gives:
\begin{equation}
    \langle A_{\boldsymbol{\eta}\boldsymbol{\nu}} \rangle = \langle \sigma_{\eta_0} \rangle \langle A_{\eta_1 \nu_1} A_{\eta_2 \nu_2} A_{\eta_3 \nu_3} \rangle
\end{equation}
where the expectation value of the center site factors out because the center is distinct from every neighbor. The joint expectation value of the three atomic densities is:
\begin{equation}
    \langle A_{\eta_1 \nu_1} A_{\eta_2 \nu_2} A_{\eta_3 \nu_3} \rangle = \sum_{j_1} \sum_{j_2} \sum_{j_3} \langle \sigma_{\eta_1 j_1} \sigma_{\eta_2 j_2} \sigma_{\eta_3 j_3} \rangle \phi_{\nu_1 j_1} \phi_{\nu_2 j_2} \phi_{\nu_3 j_3}  
\end{equation}
where we adopt the shorthand notation $\sigma_{\eta j} = \sigma_\eta (\mu_j)$ and $\phi_{\nu j} = \phi_{\nu}(\boldsymbol{r}_{i j})$.

As in the triplet case of \cref{main-sect:disordered_descriptors}, the unrestricted sums separate into a term involving only distinct sites and several terms involving self-interactions:
\begin{equation}
    \begin{aligned}
        \langle A_{\eta_1 \nu_1} A_{\eta_2 \nu_2} A_{\eta_3 \nu_3} \rangle = 
        \sum_{j_1 \neq j_2 \neq j_3} \langle \sigma_{\eta_1 j_1} \rangle \langle \sigma_{\eta_2 j_2}\rangle \langle \sigma_{\eta_3 j_3} \rangle \phi_{\nu_1 j_1} \phi_{\nu_2 j_2} \phi_{\nu_3 j_3} \\
        + \sum_{j_1 \neq j_2 = j_3 = j} \langle \sigma_{\eta_1 j_1} \rangle \langle \sigma_{\eta_2 j}  \sigma_{\eta_3 j} \rangle \phi_{\nu_1 j_1} \phi_{\nu_2 j} \phi_{\nu_3 j} \\
        + \sum_{j_2 \neq j_1 = j_3 = j} \langle \sigma_{\eta_2 j_2} \rangle \langle \sigma_{\eta_1 j}  \sigma_{\eta_3 j} \rangle \phi_{\nu_1 j} \phi_{\nu_2 j_2} \phi_{\nu_3 j} \\
        + \sum_{j_3 \neq j_1 = j_2 = j} \langle \sigma_{\eta_3 j_3} \rangle \langle \sigma_{\eta_1 j}  \sigma_{\eta_2 j} \rangle \phi_{\nu_1 j} \phi_{\nu_2 j} \phi_{\nu_3 j_3} \\
        + \sum_{j_1 = j_2 = j_3 = j} \langle \sigma_{\eta_1 j} \sigma_{\eta_2 j}  \sigma_{\eta_3 j} \rangle \phi_{\nu_1 j} \phi_{\nu_2 j} \phi_{\nu_3 j}
    \end{aligned}
\end{equation}
The constrained triple sum is removed by adding and subtracting its diagonal terms. This lifts the restriction on the triple sum and recasts some of the expectation values of the coincident sites as covariances:
\begin{equation}
    \begin{aligned}
        \langle A_{\eta_1 \nu_1} A_{\eta_2 \nu_2} A_{\eta_3 \nu_3} \rangle =
        \sum_{j_1} \sum_{j_2} \sum_{j_3} \langle \sigma_{\eta_1 j_1} \rangle \langle \sigma_{\eta_2 j_2}\rangle \langle \sigma_{\eta_3 j_3} \rangle \phi_{\nu_1 j_1} \phi_{\nu_2 j_2} \phi_{\nu_3 j_3} \\
        + \sum_{j_1 \neq j_2 = j_3 = j} \langle \sigma_{\eta_1 j_1} \rangle \text{Cov}( \sigma_{\eta_2 j}  \sigma_{\eta_3 j}) \phi_{\nu_1 j_1} \phi_{\nu_2 j} \phi_{\nu_3 j} \\
        + \sum_{j_2 \neq j_1 = j_3 = j} \langle \sigma_{\eta_2 j_2} \rangle \text{Cov}( \sigma_{\eta_1 j}  \sigma_{\eta_3 j} ) \phi_{\nu_1 j} \phi_{\nu_2 j_2} \phi_{\nu_3 j} \\
        + \sum_{j_3 \neq j_1 = j_2 = j} \langle \sigma_{\eta_3 j_3} \rangle \text{Cov}( \sigma_{\eta_1 j}  \sigma_{\eta_2 j} ) \phi_{\nu_1 j} \phi_{\nu_2 j} \phi_{\nu_3 j_3} \\
        + \sum_{j_1 = j_2 = j_3 = j} (\langle \sigma_{\eta_1 j} \sigma_{\eta_2 j}  \sigma_{\eta_3 j} \rangle - \langle \sigma_{\eta_1 j} \rangle \langle \sigma_{\eta_2 j}  \rangle \langle \sigma_{\eta_3 j} \rangle) \phi_{\nu_1 j} \phi_{\nu_2 j} \phi_{\nu_3 j}
    \end{aligned}
\end{equation}
where $\text{Cov}(\sigma_{\eta_1} \sigma_{\eta_2}) = \langle \sigma_{\eta_1}  \sigma_{\eta_2} \rangle - \langle \sigma_{\eta_1}  \rangle \langle \sigma_{\eta_2} \rangle$. The restriction on the constrained double sums of the first three self-interaction terms is lifted in the same way, and their diagonals combine with the final term to give the third joint central moment of the site basis functions:
\begin{equation}
    \begin{aligned}
        \langle A_{\eta_1 \nu_1} A_{\eta_2 \nu_2} A_{\eta_3 \nu_3} \rangle =
        \sum_{j_1} \sum_{j_2} \sum_{j_3} \langle \sigma_{\eta_1 j_1} \rangle \langle \sigma_{\eta_2 j_2}\rangle \langle \sigma_{\eta_3 j_3} \rangle \phi_{\nu_1 j_1} \phi_{\nu_2 j_2} \phi_{\nu_3 j_3} \\
        + \sum_{j_1} \sum_{j_2} \langle \sigma_{\eta_1 j_1} \rangle \text{Cov}( \sigma_{\eta_2 j_2}  \sigma_{\eta_3 j_2}) \phi_{\nu_1 j_1} \phi_{\nu_2 j_2} \phi_{\nu_3 j_2} \\
        + \sum_{j_1} \sum_{j_2} \langle \sigma_{\eta_2 j_1} \rangle \text{Cov}( \sigma_{\eta_1 j_2}  \sigma_{\eta_3 j_2} ) \phi_{\nu_1 j_2} \phi_{\nu_2 j_1} \phi_{\nu_3 j_2} \\
        + \sum_{j_1} \sum_{j_2} \langle \sigma_{\eta_3 j_1} \rangle \text{Cov}( \sigma_{\eta_1 j_2}  \sigma_{\eta_2 j_2} ) \phi_{\nu_1 j_2} \phi_{\nu_2 j_2} \phi_{\nu_3 j_1} \\
        + \sum_{j} \text{Cov}(\sigma_{\eta_1 j} \sigma_{\eta_2 j} \sigma_{\eta_3 j}) \phi_{\nu_1 j} \phi_{\nu_2 j} \phi_{\nu_3 j}
    \end{aligned}
    \label{eq:quadruplet_expectation_unconstrained}
\end{equation}
where the third joint central moment is:
\begin{equation}
    \begin{gathered}
    \text{Cov}(\sigma_{\eta_1} \sigma_{\eta_2} \sigma_{\eta_3}) = \left \langle \prod_{k = 1}^3(\sigma_{\eta_k} - \langle \sigma_{\eta_k} \rangle) \right \rangle 
    = \langle \sigma_{\eta_1} \sigma_{\eta_2} \sigma_{\eta_3} \rangle - \sum_{a} \sum_{(b, c)} \langle \sigma_{\eta_a} \rangle \langle \sigma_{\eta_b} \sigma_{\eta_c} \rangle + 2 \langle \sigma_{\eta_1} \rangle \langle \sigma_{\eta_2} \rangle  \langle \sigma_{\eta_3} \rangle 
    \end{gathered}
\end{equation}
where the double sum runs over the three ways of leaving one bond $a$ free and collapsing the pair $(b, c)$.

The expectation values of the site basis functions are composition-dependent polynomials, so they can move outside the summations. Rearranging \cref{eq:quadruplet_expectation_unconstrained} gives:
\begin{equation}
    \begin{aligned}
        \langle A_{\eta_1 \nu_1} A_{\eta_2 \nu_2} A_{\eta_3 \nu_3} \rangle = 
        \langle \sigma_{\eta_1} \rangle \langle \sigma_{\eta_2}\rangle \langle \sigma_{\eta_3} \rangle \sum_{j_1} \phi_{\nu_1 j_1} \sum_{j_2} \phi_{\nu_2 j_2} \sum_{j_3} \phi_{\nu_3 j_3} \\
        + \langle \sigma_{\eta_1} \rangle \text{Cov}( \sigma_{\eta_2}  \sigma_{\eta_3}) \sum_{j_1} \phi_{\nu_1 j_1} \sum_{j_2} \phi_{\nu_2 j_2} \phi_{\nu_3 j_2} \\
        + \langle \sigma_{\eta_2} \rangle \text{Cov}( \sigma_{\eta_1}  \sigma_{\eta_3} ) \sum_{j_1} \phi_{\nu_2 j_1} \sum_{j_2} \phi_{\nu_1 j_2} \phi_{\nu_3 j_2} \\
        + \langle \sigma_{\eta_3} \rangle \text{Cov}( \sigma_{\eta_1}  \sigma_{\eta_2} ) \sum_{j_1} \phi_{\nu_3 j_1} \sum_{j_2}  \phi_{\nu_1 j_2} \phi_{\nu_2 j_2} \\
        + \text{Cov}(\sigma_{\eta_1} \sigma_{\eta_2} \sigma_{\eta_3}) \sum_{j} \phi_{\nu_1 j} \phi_{\nu_2 j} \phi_{\nu_3 j}
    \end{aligned}
\end{equation}
The average of $A_{\boldsymbol{\eta}\boldsymbol{\nu}}$ finally becomes:
\begin{equation}
    \begin{aligned}
        \langle A_{\boldsymbol{\eta}\boldsymbol{\nu}} \rangle = \langle \sigma_{\eta_0} \rangle \Big [ \langle \sigma_{\eta_1} \rangle \langle \sigma_{\eta_2}\rangle \langle \sigma_{\eta_3 } \rangle  A_{\nu_1} A_{\nu_2} A_{\nu_3} 
        + \sum_a \sum_{(b, c)} \langle \sigma_{\eta_a} \rangle \text{Cov}( \sigma_{\eta_b}  \sigma_{\eta_c}) A_{\nu_a} A_{(\nu_b \nu_c)}^{\text{self}} 
        + \text{Cov}(\sigma_{\eta_1} \sigma_{\eta_2} \sigma_{\eta_3}) A_{(\nu_1 \nu_2 \nu_3)}^{\text{self}} \Big ]
    \end{aligned}
    \label{eq:average_quadruplet_A}
\end{equation}
where the superscript \textit{self} marks atomic densities evaluated on self-interacting bond functions, namely single-bond functions formed as products of two or three bond functions evaluated on the same bond. The symmetrized average $B$-basis function, $\langle B_{\boldsymbol{\eta}\boldsymbol{\nu}} \rangle$, follows from an application of the Clebsch-Gordan iteration to \cref{eq:average_quadruplet_A}.

\section{Expressions for the self-interaction descriptors} \label{sect:self_interaction}
This section reports analytical expressions for the symmetrized triplet and quadruplet self-interaction descriptors. We work throughout with complex spherical harmonics as defined in Varshalovich \textit{et al.} \cite{varshalovich_quantum_1988}. The Legendre polynomials within the spherical harmonics include the Condon-Shortley phase. The prefactor $N_{l_1 l_2}^L$ in \cref{main-eq:inverse_cg_series} takes the form \cite{varshalovich_quantum_1988}:
\begin{equation}
    N_{l_1 l_2}^L = \sqrt{\frac{(2 l_1 + 1)(2 l_2 + 1)}{4 \pi (2 L +1)}}
    \label{eq:prefactor_inverse_cg_series}
\end{equation}
Substituting \cref{eq:prefactor_inverse_cg_series} in \cref{main-eq:triplets_two_bonds}, together with $C_{l 0 l 0}^{0 0} = (-1)^l / \sqrt{2l +1}$ \cite{varshalovich_quantum_1988}, gives the triplet self-interaction:
\begin{equation}
    B_{n_1 n_2 l}^{\text{self}} =  (-1)^l \sqrt{\frac{2l + 1}{4 \pi}} B_{(n_1 n_2)0}
\end{equation}
A quadruplet admits three self-interacting clusters with two coincident bonds and one cluster with all three bonds collapsed. Substituting \cref{eq:prefactor_inverse_cg_series} in \cref{main-eq:quardruplets_two_bonds} gives one of the quadruplet self-interactions with two coincident bonds:
\begin{equation}
    B_{n_1 (n_2 n_3) l_1 (l_2 l_3)}^{\text{self}} = \sqrt{\frac{(2 l_2 + 1)(2 l_3 + 1)}{4 \pi (2 l_1 +1)}} C_{l_2 0 l_3 0}^{l_1 0} B_{n_1 (n_2 n_3) l_1}
\end{equation}
The remaining two self-interaction descriptors follow from the symmetry of the invariant combination of three spherical harmonics. As mentioned in \cref{sect:quadruplet_descriptors}, the symmetrized quadruplet $B$-basis functions emerge from an application of the Clebsch-Gordan iteration depicted in \cref{main-fig:cg_iteration}(b) to the atomic densities. Inspecting \cref{eq:average_quadruplet_A}, we arrive at the generic symmetrized self-interaction descriptor with two coincident bonds:
\begin{equation}
    B_{n_a (n_b n_c) l_a (l_b l_c)}^{\text{self}} = \sum_{m_a m_b m_c} C_{l_a -m_a l_a m_a}^{00} C_{l_b m_b l_c m_c}^{l_a -m_a} A_{n_a l_a} A_{(n_b n_c) (l_b l_c)}^{\text{self}} = \sum_{m_a m_b m_c} \left ( \begin{matrix} 
    l_b \quad l_c \quad l_a \\
    m_b \quad m_c \quad m_a
    \end{matrix}
    \right ) A_{n_a l_a} A_{(n_b n_c) (l_b l_c)}^{\text{self}}
    \label{eq:wigner_quardruplets_two_bonds}
\end{equation}

In \cref{eq:wigner_quardruplets_two_bonds}, the two-step angular contraction is compactly represented by a Wigner-$3j$ symbol \cite{edmonds_angular_1960, varshalovich_quantum_1988}. Importantly, no phase factor appears because the sum of the $l$ quantum numbers is even. Because the Wigner-$3j$ symbol is symmetric under even permutations \cite{edmonds_angular_1960, varshalovich_quantum_1988}, the various self-interaction descriptors arise by permuting the $n$, $l$, and $m$ indices in \cref{eq:wigner_quardruplets_two_bonds}. The remaining two self-interaction descriptors are thus:
\begin{equation}
    \begin{gathered}
        B_{n_2 (n_3 n_1) l_2 (l_3 l_1)}^{\text{self}} = N_{l_3 l_1}^{l_2} C_{l_3 0 l_1 0}^{l_2 0} B_{n_2 (n_3 n_1) l_2} = (-1)^{l_3} \sqrt{\frac{(2 l_3 + 1)}{4 \pi}} C_{l_2 0 l_3 0}^{l_1 0} B_{n_2 (n_3 n_1) l_2} \\
        B_{n_3 (n_1 n_2) l_3 (l_1 l_2)}^{\text{self}} =  N_{l_1 l_2}^{l_3} C_{l_1 0 l_2 0}^{l_3 0} B_{n_3 (n_1 n_2) l_3} = (-1)^{l_2} \sqrt{\frac{(2 l_2 + 1)}{4 \pi}} C_{l_2 0 l_3 0}^{l_1 0} B_{n_3 (n_1 n_2) l_3}
    \end{gathered}
\end{equation}
where we made use of $C_{l_2 0 l_3 0}^{l_1 0} = (-1)^{l_3} \sqrt{\frac{(2 l_1 + 1)}{(2 l_2 + 1)}} C_{l_3 0 l_1 0}^{l_2 0} = (-1)^{l_2} \sqrt{\frac{(2 l_1 + 1)}{(2 l_3 + 1)}} C_{l_1 0 l_2 0}^{l_3 0}$ \cite{varshalovich_quantum_1988} to simplify the expressions.

When all three bonds collapse into one, the quadruplet self-interaction reduces to \cref{main-eq:quardruplets_three_bonds}. Substituting for $N_{l_1 l_1}^{0}$, $N_{l_2 l_3}^{l_1}$, and $C_{l_1 0 l_1 0}^{0 0}$ then produces:
\begin{equation}
    B_{(n_1 n_2 n_3) (l_1 l_2 l_3)}^{\text{self}} = (-1)^{l_1} \frac{\sqrt{(2 l_2 + 1)(2 l_3 + 1)}}{4 \pi} C_{l_2 0 l_3 0}^{l_1 0} B_{(n_1 n_2 n_3) 0}
\end{equation}

\section{Density functional theory calculations} \label{sect:dft_calculations}
The density functional theory (DFT) calculations for Mg--Nd used a plane-wave cutoff of 500\,eV, a $\Gamma$-centered $k$-point mesh with a spacing no larger than 0.02\,\AA$^{-1}$, and a second-order Methfessel-Paxton smearing of 0.2\,eV. Those for Cr--W and Mo--Nb used a plane-wave cutoff of 550\,eV, a $\Gamma$-centered $k$-point mesh with a spacing no larger than 0.018\,\AA$^{-1}$, and a second-order Methfessel-Paxton smearing of 0.1\,eV. The Cr--W and Mo--Nb calculations were spin-polarized, with the magnetic moments treated within collinear magnetism. \Cref{tab:pseudopotentials} lists the valence electron configuration of the projector-augmented-wave (PAW) pseudopotential used for each element. Part of the data were taken from earlier work \cite{piersante_machine_2026, lee_modeling_2026}.

\begin{table}[!ht]
    \begin{tabular}{| c | c | c |} \hline
    Element & Valence configuration & Version \\ \hline
    Mg & 2p$^6$3s$^2$ & \verb|Mg_pv| 13Apr2007 \\
    Nd & 5s$^2$5p$^6$4f$^1$6s$^2$ & \verb|Nd_3| 06Sep2000 \\
    Cr & 3p$^6$3d$^5$4s$^1$ & \verb|Cr_pv| 02Aug2007 \\
    W & 5s$^2$5p$^6$5d$^5$6s$^1$ & \verb|W_sv| 04Sep2015 \\
    Mo & 4s$^2$4p$^6$4d$^5$5s$^1$ & \verb|Mo_sv| 02Feb2006 \\
    Nb & 4s$^2$4p$^6$4d$^4$5s$^1$ & \verb|Nb_sv| 25May2007 \\ \hline
    \end{tabular}
    \caption{Valence electron configuration of the projector-augmented-wave (PAW) pseudopotential used for each element. The version column gives the name and release date of the VASP PAW dataset.}
    \label{tab:pseudopotentials}
\end{table}

\section{Datasets} \label{sect:datasets}
\Cref{tab:MgNd_database} breaks down the reference Mg--Nd dataset used to study how the global errors converge with the number of basis functions and with the training set size. \Cref{tab:MgNd_point_training_set,tab:MgNd_point_validation_set} summarize the datasets used in the study of point defects, while \cref{tab:MoNb_training_set,tab:CrW_training_set} detail those used for the convergence of the mixing enthalpy.

\begin{table}[!ht]
    \begin{minipage}[t]{0.28\linewidth}
        \centering
        \small
        \begin{tabular}{|c|c|}\hline
            Structure type & Count \\ \hline
            Orderings on HCP & 1794 \\
            Orderings on BCC & 2098 \\
            Orderings on C15 & 836 \\
            Mg--Nd phases &  18 \\ \hline
            Total & 4746 \\ \hline
        \end{tabular}
        \caption{Mg--Nd reference dataset for the study of error convergence trends (\cref{main-sect:error_metrics_point_defects}).}
        \label{tab:MgNd_database}
    \end{minipage}\hfill
    \begin{minipage}[t]{0.38\linewidth}
        \centering
        \small
        \begin{tabular}{|c|c|}\hline
            Structure type & Count \\ \hline
            HCP/BCC Mg & 6 \\
            Vacancy in HCP/BCC Mg & 6 \\
            GSF/SSF/GB in HCP Mg & 91 \\
            Nd point/pair in HCP Mg & 28 \\
            $\beta^\prime$/$\beta^{\prime\prime}$/$\beta_1$/C15/B2 phases & 11 \\
            HCP/DHCP Nd & 3 \\ \hline
            Total & 145 \\ \hline
        \end{tabular}
        \caption{Training dataset used in the study of Mg--Nd point defects (\cref{main-sect:error_metrics_point_defects}). The planar-fault structures comprise generalized stacking faults (GSF), stable stacking faults (SSF), and grain boundaries (GB).}
        \label{tab:MgNd_point_training_set}
    \end{minipage}\hfill
    \begin{minipage}[t]{0.30\linewidth}
        \centering
        \small
        \begin{tabular}{|c|c|}\hline
            Structure type & Count \\ \hline
            Nd point in SSF & 674 \\
            Vacancies in Mg--Nd phases & 435 \\ \hline
            Total & 1109 \\ \hline
        \end{tabular}
        \caption{Validation dataset of Nd point defects close to planar faults and of vacancies in the $\beta^\prime$, $\beta^{\prime\prime}$, $\beta^{\prime\prime\prime}$, $\beta_1$, and C15 phases (\cref{main-sect:error_metrics_point_defects}).}
        \label{tab:MgNd_point_validation_set}
    \end{minipage}
\end{table}

\begin{table}[!ht]
    \begin{minipage}[t]{0.48\linewidth}
        \centering
        \small
        \begin{tabular}{|c|c|}\hline
            Structure type & Count \\ \hline
            Orderings on BCC & 137 \\
            Volumetric perturbations & 22 \\ \hline
            Total & 159 \\ \hline
        \end{tabular}
        \caption{Mo--Nb training dataset (\cref{main-sect:random_alloy}).}
        \label{tab:MoNb_training_set}
    \end{minipage}\hfill
    \begin{minipage}[t]{0.48\linewidth}
        \centering
        \small
        \begin{tabular}{|c|c|}\hline
            Structure type & Count \\ \hline
            Orderings on BCC & 135 \\
            Volumetric perturbations & 22 \\ \hline
            Total & 157 \\ \hline
        \end{tabular}
        \caption{Cr--W training dataset (\cref{main-sect:random_alloy}).}
        \label{tab:CrW_training_set}
    \end{minipage}
\end{table}

\section{Interatomic potential configurations} \label{sect:potential_config}
\Cref{tab:potential_settings} lists the radial and angular basis set parameters of the ACE potentials fitted in this work, together with the cutoff radius used for each system. The size of the basis was limited to prevent overfitting to the small training sets of \cref{main-sect:error_metrics_point_defects,main-sect:random_alloy}, and was selected from the train and test errors plotted against the number of basis functions in \cref{main-fig:basis_functions}.

\begin{table}[!ht]
    \centering
    \begin{tabular}{|c | c  c  c | c | c |}
        \hline
        \multirow{2}{*}{Potential} & \multicolumn{3}{|c|}{$n$/$l$} & \multirow{2}{*}{Total} & \multirow{2}{*}{$r_\mathrm{cut}$ (\AA)} \\
        ~ & Pair & Triplet & Quadruplet & ~ & ~ \\ \hline
        Mg--Nd/P & 11/0 & -- & -- & 44 & 7.8 \\
        Mg--Nd/T & 11/0 & 4/4 & -- & 404 & 7.8 \\
        Mg--Nd/Q & 11/0 & 4/4 & 2/2 & 724 & 7.8 \\
        Mo--Nb/P & 11/0 & -- & -- & 44 & 7.0 \\
        Mo--Nb/T & 11/0 & 2/2 & -- & 104 & 7.0 \\
        Cr--W/P & 11/0 & -- & -- & 44 & 7.0 \\
        Cr--W/T & 11/0 & 2/2 & -- & 104 & 7.0 \\ \hline
    \end{tabular}
    \caption{Basis set parameters of the ACE potentials fitted in this work. Each row is labeled by the alloy system and by the truncation of the cluster basis, with P, T, and Q denoting the pair, triplet, and quadruplet potentials. The $n$/$l$ columns give the maximum degree of the radial and angular basis functions at each body order, Total the resulting number of $B$-basis functions, and $r_\mathrm{cut}$ the cutoff radius.}
    \label{tab:potential_settings}
\end{table}

\section{Supplementary results}

\subsection{Point defects}
\begin{figure}[!ht]
    \centering
    \includegraphics[width=0.6\linewidth]{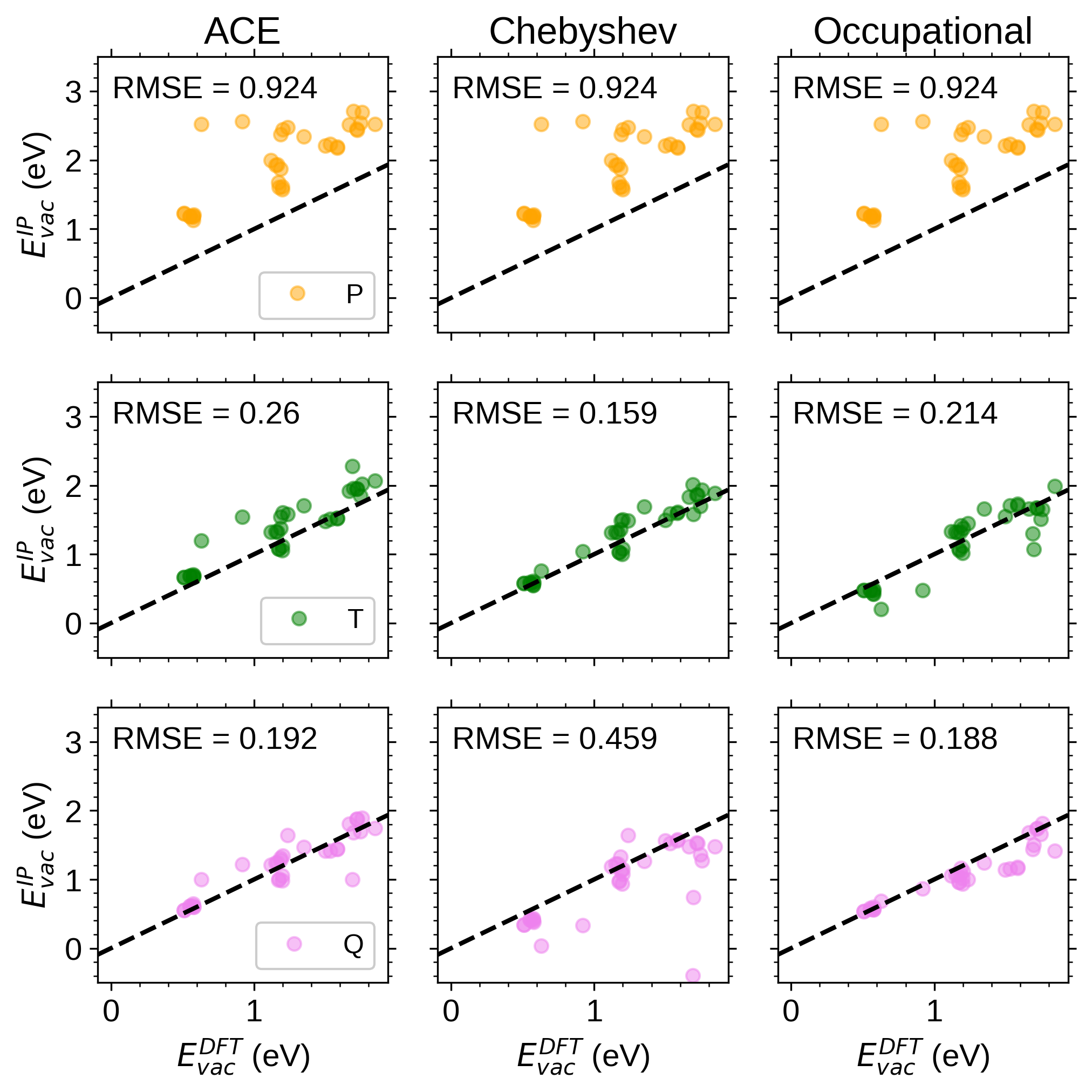}
    \caption{Parity plots of the vacancy formation energies predicted by the interatomic potentials against the DFT reference. The three columns correspond to the conventional ACE, Chebyshev, and occupational site bases, and the rows to the pair (P), triplet (T), and quadruplet (Q) potentials. The dashed line marks perfect agreement, and each panel is annotated with its RMSE in eV.}
    \label{fig:vacancy_parity_plots}
\end{figure}

The ACE models of \cref{main-sect:error_metrics_point_defects} were fitted by Ridge regression with leave-one-out cross validation.
\Cref{tab:point_defects_training_energy_errors,tab:point_defects_training_force_errors} report the resulting training errors on energies and forces. \Cref{fig:vacancy_parity_plots} compares the vacancy formation energies predicted by these potentials against the DFT reference, resolved by chemical site basis and by body order.

\begin{figure}[!ht]
    \centering
    \subfloat[][Prismatic \label{fig:Nd_prismatic}]{\includegraphics[width=0.5\linewidth]{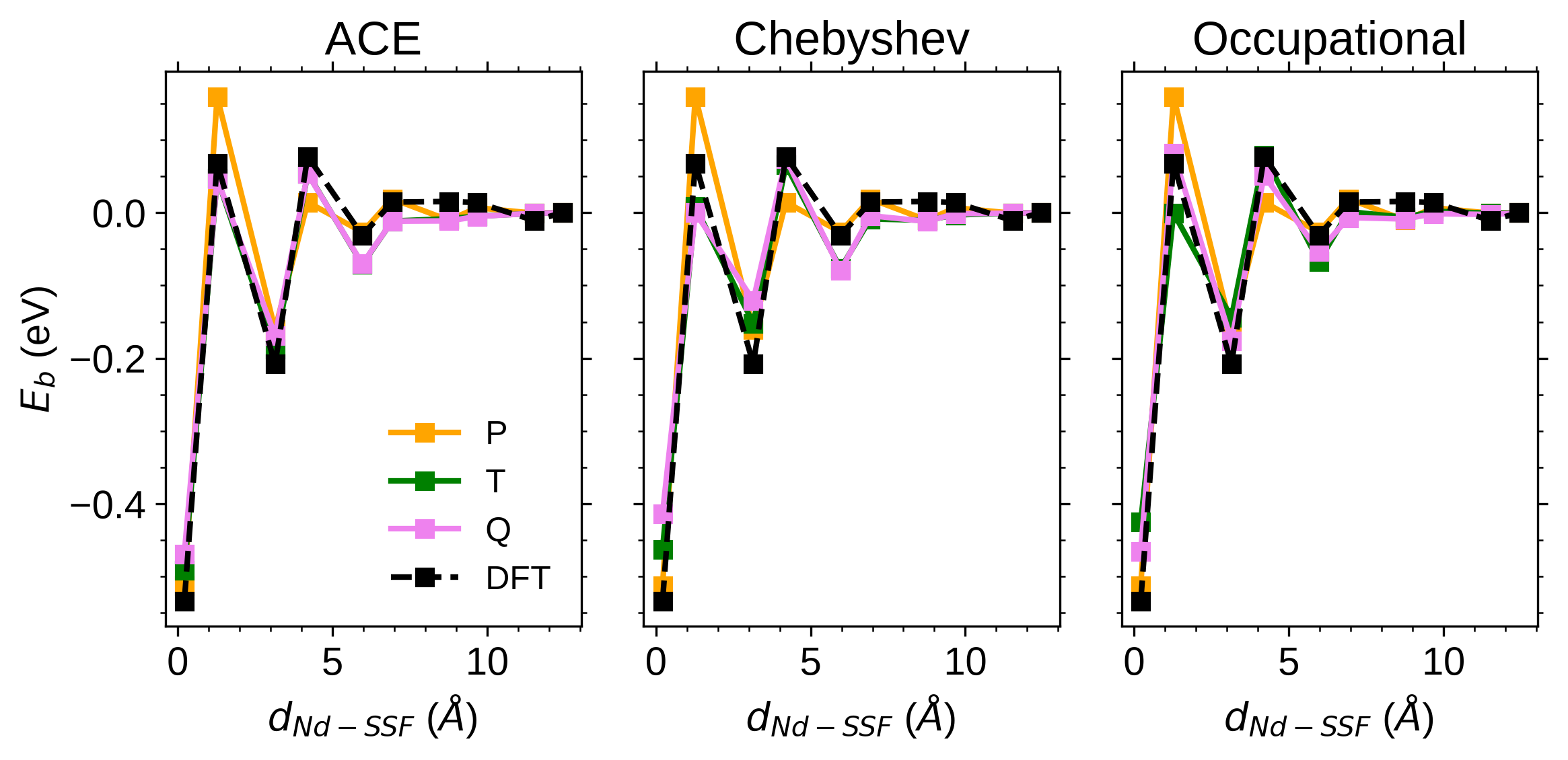}}
    \subfloat[][Pyramidal II \label{fig:Nd_pyramid2}]{\includegraphics[width=0.5\linewidth]{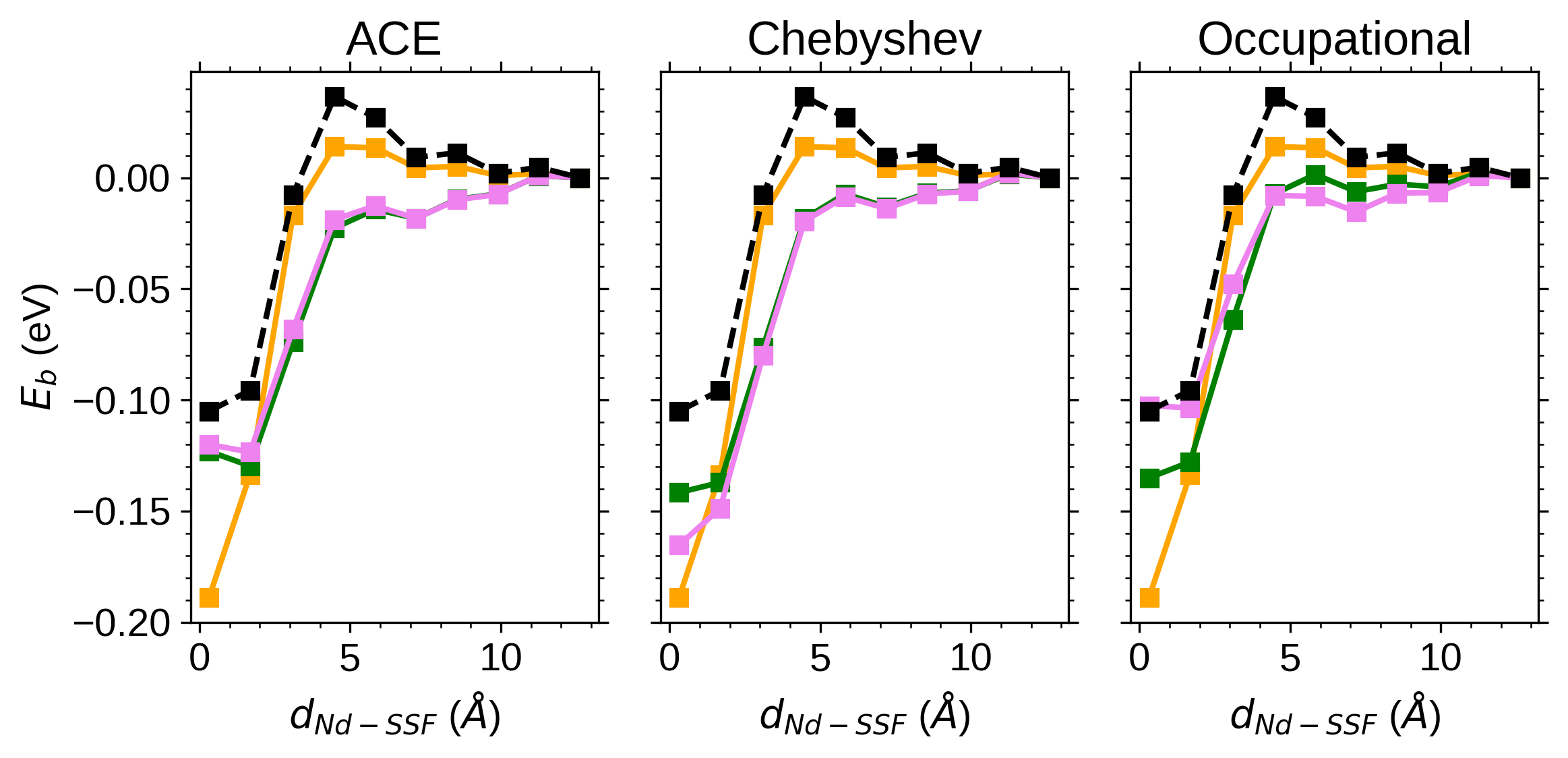}}
    \caption{Binding energy of a Nd solute to a prismatic (a) and a pyramidal II (b) stable stacking fault in HCP Mg, as a function of the distance between the solute and the fault. Within each panel, the three columns correspond to the conventional ACE, Chebyshev, and occupational site bases, and P, T, and Q denote the pair, triplet, and quadruplet potentials compared against the DFT reference. Negative energies indicate binding.}
\end{figure}

\Cref{fig:Nd_prismatic,fig:Nd_pyramid2} show the binding energy of a Nd solute to the prismatic and pyramidal II SSFs in HCP Mg. As for the pyramidal I and basal faults of the main text, the three bases give comparable results for the prismatic SSF. For the pyramidal II SSF, only the quadruplet potential built on the occupational basis reproduces both the magnitude and the shape of the binding energy profile, while the ACE and Chebyshev bases predict excessively negative binding energies near the fault.

\begin{table}[!ht]
    \begin{minipage}[t]{0.45\linewidth}
        \centering
        \small
        \begin{tabular}{|c|c|c|c|}
        \hline
            RMSE (meV/atom) & 11/0 & 11/4-0/4 & 11/4/2-0/4/2 \\ \hline
            ACE & 4.9 & 0.5 & 0.5 \\
            Chebyshev & 4.9 & 0.4 & 2.4\\
            Occupational & 4.9 & 0.5 & 0.5 \\ \hline
        \end{tabular}
        \caption{Energy training errors for the ACE fits of \cref{main-sect:error_metrics_point_defects}.}
        \label{tab:point_defects_training_energy_errors}
    \end{minipage}\hfill
    \begin{minipage}[t]{0.45\linewidth}
        \centering
        \small
        \begin{tabular}{|c|c|c|c|}
        \hline
            RMSE (meV/\AA) & 11/0 & 11/4-0/4 & 11/4/2-0/4/2 \\ \hline
            ACE & 17.0 & 4.7 & 4.5 \\
            Chebyshev & 17.0 & 4.4 & 3.9 \\
            Occupational & 17.0 & 4.3 & 5.1 \\ \hline
        \end{tabular}
        \caption{Force training errors for the ACE fits of \cref{main-sect:error_metrics_point_defects}.}
        \label{tab:point_defects_training_force_errors}
    \end{minipage}
\end{table}

\begin{figure}[!ht]
    \centering
    \subfloat[][\label{fig:MoNb_hulls} Mo--Nb]{\includegraphics[width=0.5\linewidth]{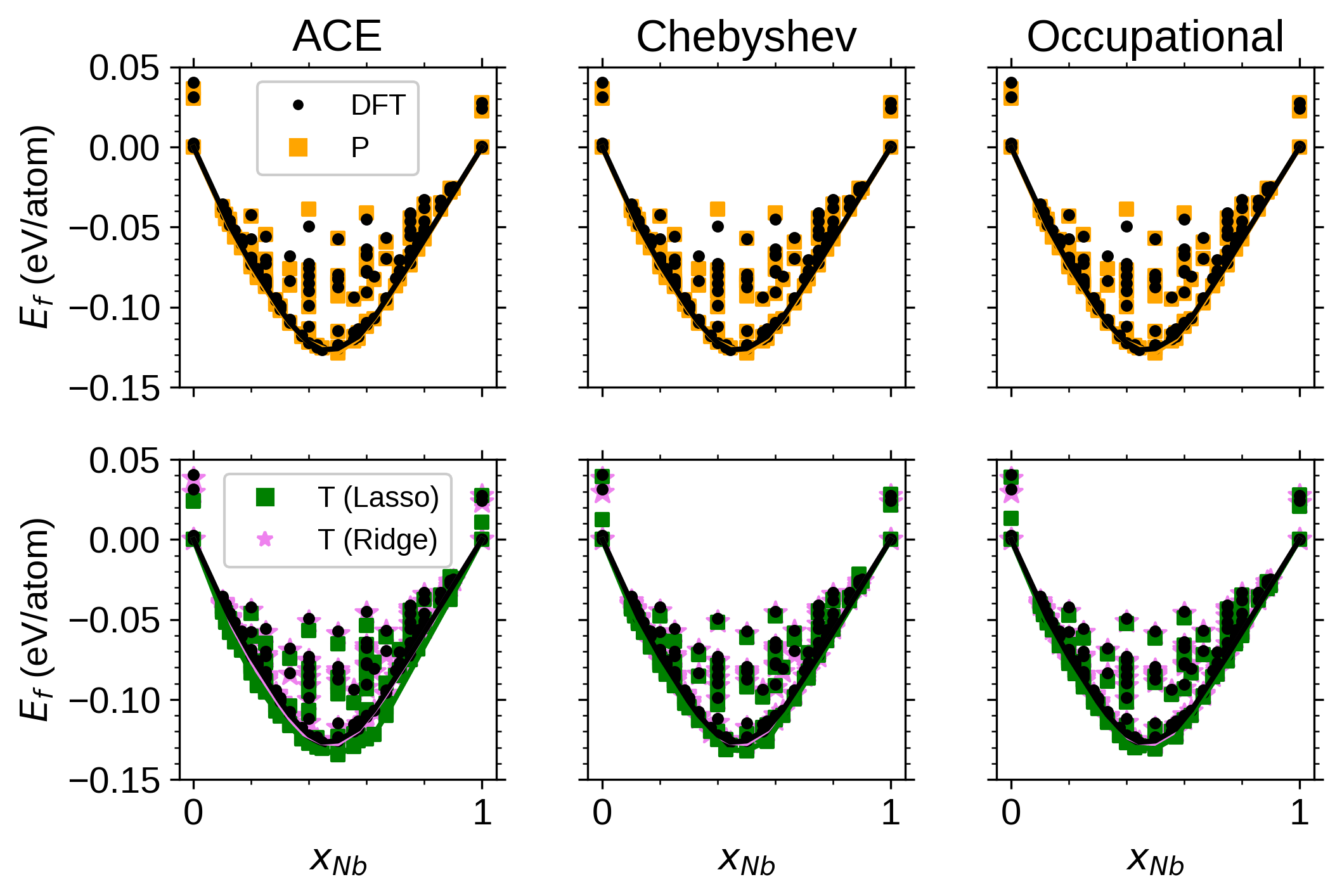}}
    \subfloat[][\label{fig:CrW_hulls} Cr--W]{\includegraphics[width=0.5\linewidth]{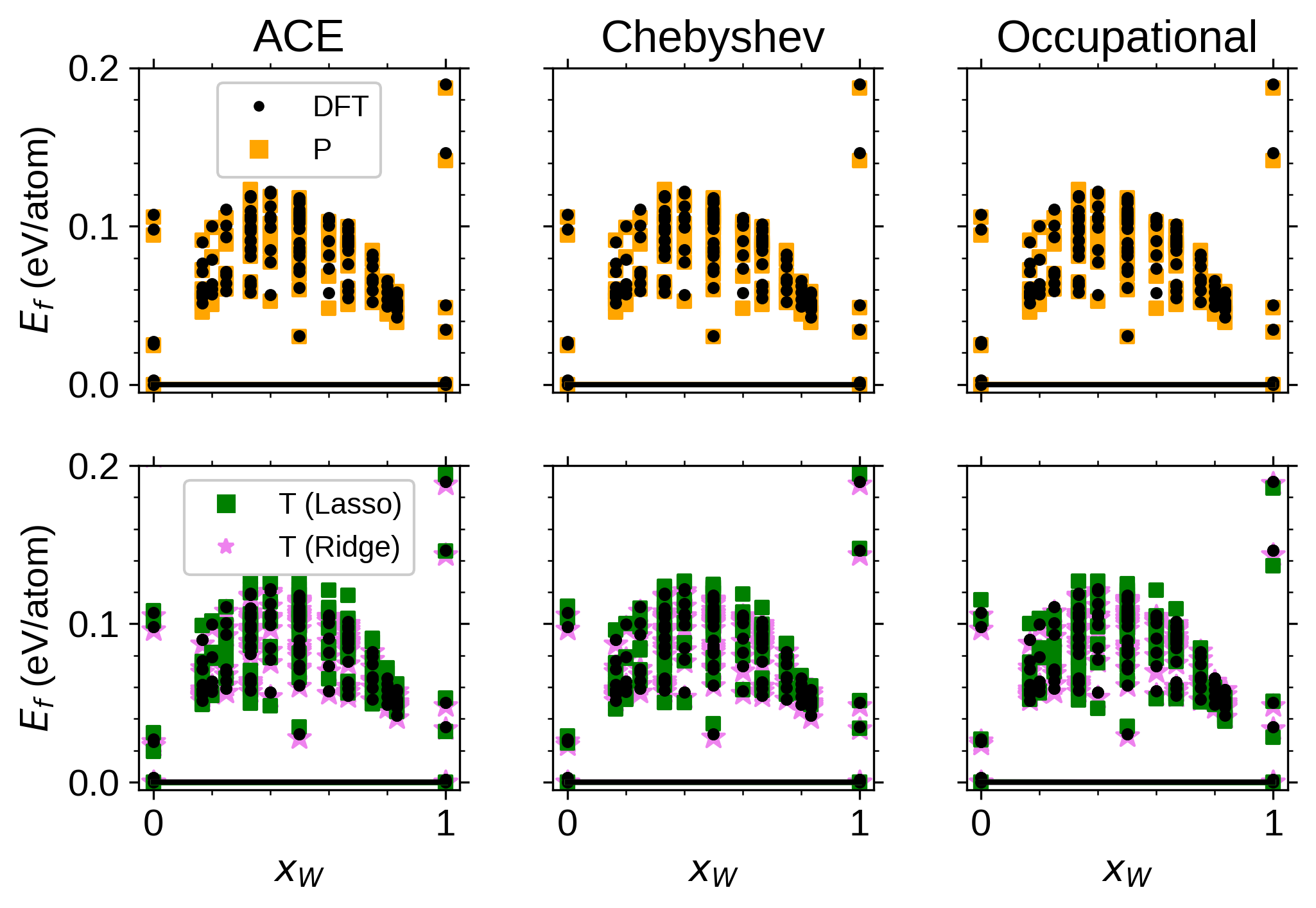}}
    \caption{Formation energies and convex hulls predicted by the ACE models for (a) Mo--Nb and (b) Cr--W. Within each panel, the three columns correspond to the conventional ACE, Chebyshev, and occupational site bases. The upper row shows the pair potential and the lower row the triplet potential fitted with either Lasso or Ridge regression, both compared against the DFT reference.}
\end{figure}

\subsection{Random alloys}
The mixing enthalpy curves of \cref{main-fig:MoNb_mixing_enthalpy,main-fig:CrW_mixing_enthalpy} were obtained from ACE models fitted by either Ridge regression or Lasso.
\Cref{fig:MoNb_hulls,fig:CrW_hulls} show the corresponding formation energy convex hulls for each body order and regression method. The hulls are qualitatively similar across the three bases and both regression methods. \Cref{tab:MoNb_rmse,tab:CrW_rmse} show, however, that the triplet potentials fitted with Lasso reproduce the DFT formation energies less accurately than those fitted with Ridge regression. The disordered phase shows the opposite trend. With the Chebyshev and occupational bases, the Lasso fits reproduce the mixing enthalpies of \cref{main-fig:MoNb_mixing_enthalpy,main-fig:CrW_mixing_enthalpy} more accurately than the Ridge fits.

For the convergence of the mixing enthalpy with training set size in \cref{main-fig:MoNb_mixing_enthalpy_convergence,main-fig:CrW_mixing_enthalpy_convergence}, the training set was grown in 5\% increments from 15\% to 90\% of the full dataset, with a final fit on the complete dataset for reference. These fits used Lasso with leave-one-out cross validation.

\begin{figure}[!ht]
    \centering
    \subfloat[][\label{fig:learning_curves_monb_crw_lasso}]{\includegraphics[width=0.5\linewidth]{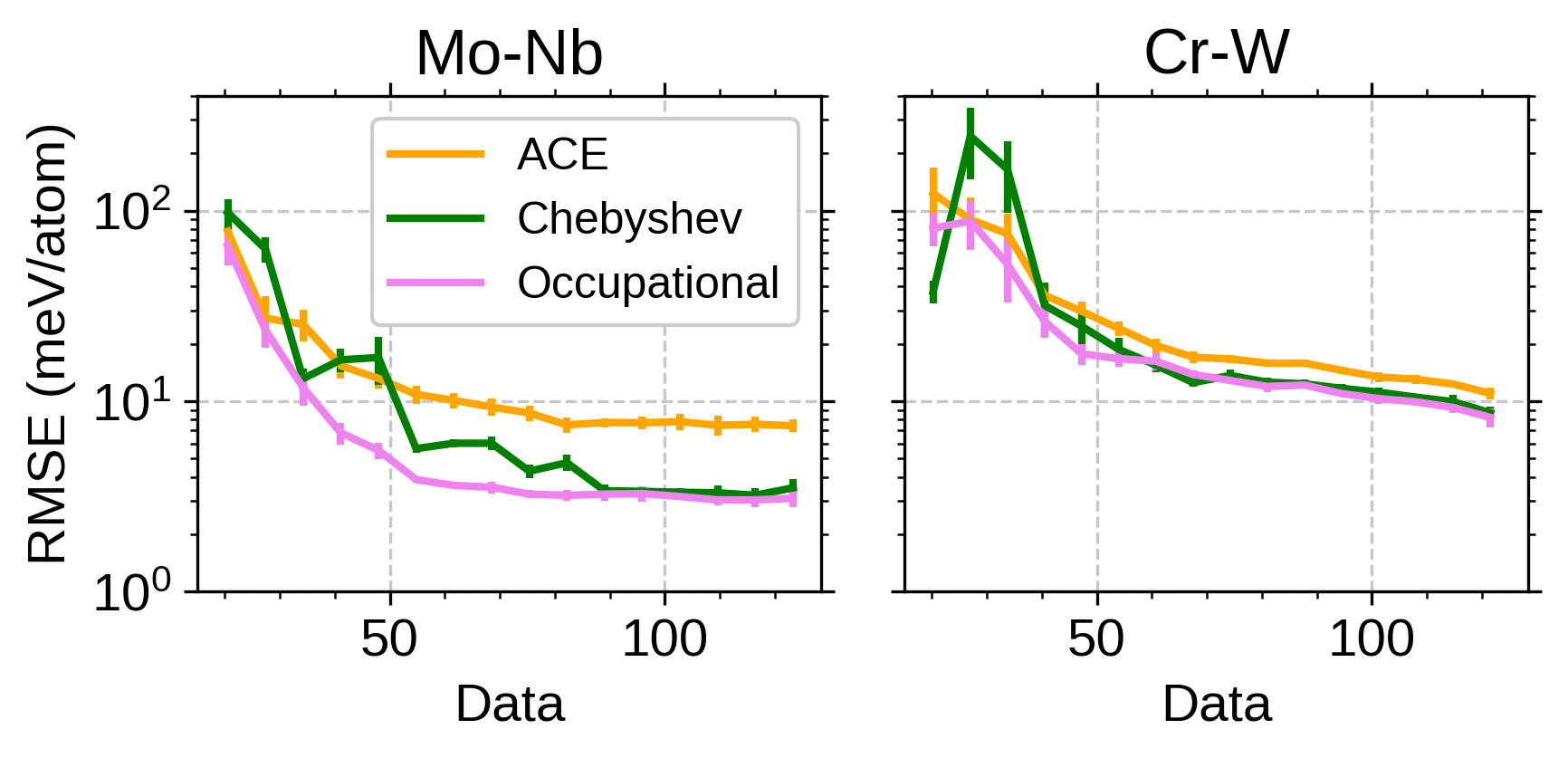}}
    \subfloat[][\label{fig:variance_convergence_lasso}]{\includegraphics[width=0.5\linewidth]{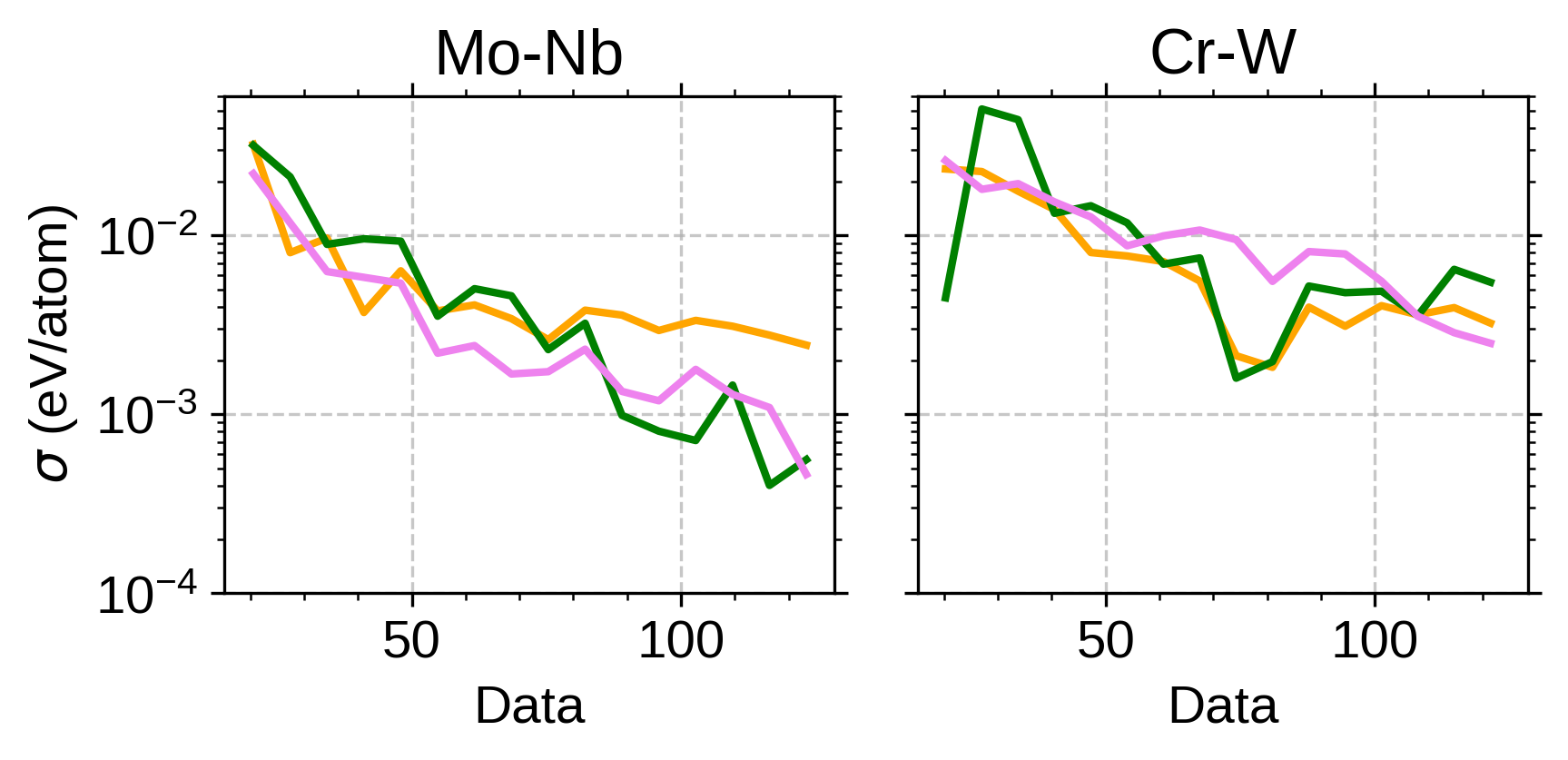}}
    \caption{
    Convergence of the Lasso-fitted potentials with training set size. (a) Learning curves of the test error, evaluated over all the left-out data. (b) Standard deviation of the mixing enthalpy at equiatomic composition, evaluated over the ten fitted average-atom potentials.
    }
\end{figure}

\Cref{fig:learning_curves_monb_crw_lasso} shows the learning curves of these Mo--Nb and Cr--W potentials, with the test error evaluated on the left-out data. \Cref{fig:variance_convergence_lasso} shows how the standard deviation of the predicted mixing enthalpy evolves with training set size, evaluated at equiatomic composition over the ten fitted average-atom potentials.

\begin{table}[!ht]
    \begin{minipage}[t]{0.48\linewidth}
        \centering
        \small
            \begin{tabular}{|c|c|c|c|}
            \hline
                RMSE (meV/atom) & 11/0 & 11/2-0/2 (Ridge) & 11/2-0/2 (Lasso) \\ \hline
                ACE & 2.12 & 0.78 & 5.98 \\ 
                Chebyshev & 2.12 & 0.80 & 3.91 \\ 
                Occupational & 2.12 & 0.87 & 3.67 \\ \hline
            \end{tabular}
            \caption{Root mean squared error of the Mo--Nb formation energies shown in \cref{fig:MoNb_hulls}, relative to the DFT reference. The columns are labeled by the radial and angular basis set sizes of \cref{tab:potential_settings} and, for the triplet fits, by the regression method.}
            \label{tab:MoNb_rmse}
    \end{minipage}\hfill
    \begin{minipage}[t]{0.48\linewidth}
        \centering
        \small
            \begin{tabular}{|c|c|c|c|}
            \hline
                RMSE (meV/atom) & 11/0 & 11/2-0/2 (Ridge) & 11/2-0/2 (Lasso) \\ \hline
                ACE & 2.30 & 0.89 & 7.13 \\ 
                Chebyshev & 2.30 & 0.91 & 5.88 \\ 
                Occupational & 2.30 & 1.00 & 6.60 \\ \hline
            \end{tabular}
            \caption{Root mean squared error of the Cr--W formation energies shown in \cref{fig:CrW_hulls}, relative to the DFT reference. The columns are labeled by the radial and angular basis set sizes of \cref{tab:potential_settings} and, for the triplet fits, by the regression method.}
            \label{tab:CrW_rmse}
    \end{minipage}
\end{table}

\bibliographystyle{unsrt}
\bibliography{bibliography}